\documentclass{llncs}
\usepackage{graphicx}
\usepackage{subcaption} 
\usepackage{float}%
\usepackage{color}
\usepackage{tikz}
\usetikzlibrary{shapes,arrows,patterns,positioning} 
\usepackage{caption}
\usepackage[normalem]{ulem}

\usepackage{verbatim}
\usepackage{multirow}
\usepackage[ruled,linesnumbered]{algorithm2e}
\usepackage{amssymb}
\usepackage{amsmath}
\usepackage{empheq}

\usepackage{breakcites} 

\usepackage{xspace,tcolorbox}
\usepackage[hyphens]{url}
\usepackage[hidelinks]{hyperref}
\hypersetup{breaklinks=true}

\usepackage[utf8]{inputenc}
\usepackage{textcomp}

\tikzstyle{decision} = [diamond, draw, 
    text width=4.5em, text badly centered, node distance=3cm, inner sep=0pt]
\tikzstyle{block} = [rectangle, draw, 
    text width=5em, text centered, rounded corners, minimum height=4em]
\tikzstyle{line} = [draw, -latex']
\tikzstyle{cloud} = [draw, ellipse,fill=red!20, node distance=3cm,
    minimum height=2em]
    \newenvironment{cppalgorithm}[1][htb]
  {\renewcommand{\algorithmcfname}{Program}
   \begin{algorithm}[#1]%
  }{\end{algorithm}}

\newcommand{\todo}[1]{{\color{red}#1}}

\newcommand{\REM}[1]{} 
\REM{ multiline
comment } 

\usepackage[formats]{listings}
\usepackage{xcolor}

\begin{document}


\title{Taming the War in Memory: A Resilient Mitigation Strategy Against Memory Safety Attacks in CPS}

\author{Eyasu Getahun Chekole\inst{1,2} \and Unnikrishnan Cheramangalath\inst{1} \and Sudipta Chattopadhyay \inst{1} \and Mart\'in Ochoa\inst{1,3} \and Guo Huaqun\inst{2}}
\authorrunning{Chekole et al.} 

\institute{Singapore University of Technology and Design, Singapore, Singapore\\
\and Institute for Infocomm Research (I$^2$R), Singapore, Singapore\\
\and Department of Applied Mathematics and Computer Science, Universidad del Rosario, Bogot\'a, Colombia}

%

\maketitle

\thispagestyle{plain} 

%

\begin{abstract}
Memory-safety attacks have been one of the most critical threats against computing systems. Although a wide-range of defense techniques have been developed against these attacks, the existing mitigation strategies have several limitations. In particular, most of the 
existing mitigation approaches 
are based on aborting or restarting the victim program when a memory-safety attack is detected, thus making the system unavailable. 
This might not be acceptable in systems with stringent timing constraints, such as cyber-physical systems (CPS), since the system unavailability leaves the control system in an unsafe state. To address this problem, we propose CIMA -- a resilient and light-weight mitigation technique that prevents invalid memory accesses at runtime. 
CIMA manipulates the compiler-generated control-flow graph to automatically detect and bypass unsafe memory accesses at runtime, thereby mitigating memory-safety attacks along the process. An appealing feature of CIMA is that it also ensures system availability and resilience of the CPS 
even under the presence of memory-safety attacks. To this end, we design our experimental setup based on a realistic Secure Water Treatment (SWaT) and Secure Urban Transportation System (SecUTS) testbeds and evaluate the effectiveness and the efficiency of our approach. 
The experimental results reveal that CIMA handles memory-safety attacks effectively with low overhead. Moreover, it meets the real-time constraints and physical-state resiliency of the CPS under test. 

\end{abstract}

\section{Introduction}\label{sec:introduction}


%

 

Software systems with stringent real-time constraints are often written in C/C++ since they aid in generating efficient program binaries. 
However, since memory management is handled manually and also the lack of bounds checking in C/C++, programs written in such languages often suffer from memory-safety vulnerabilities, such as buffer over/underflows, dangling pointers, double-free errors and memory leaks. Due to the difficulty in discovering these vulnerabilities, they might often slip into production run. 
This leads to memory corruptions and runtime crashes even in deployed systems.
In the worst case, memory-safety vulnerabilities can be exploited by a class of cyber attacks, which we refer to as \emph{memory-safety attacks}~\cite{sok,memory_safety_attacks_survey}. Memory-safety attacks, such as code injection \cite{code_injection_survey,code_injection_attacks} and code reuse \cite{code_reuse_attacks1,code_reuse_attacks2,brop}, can cause devastating effects by compromising vulnerable programs in any computing system. These attacks can hijack or subvert specific operations of a system or take over the entire system, in general. A detailed account of such attacks is provided in existing works~\cite{brop,eyasu_cybericps,eyasu_essos}. 

Given a system vulnerable to memory-safety attacks, its runtime behavior will depend on the type of memory accesses being made, among others. For instance, accesses to an array element beyond the imposed length of the array (buffer overflows) could be exploited to overwrite the return address of a function. 
However, a safety check generated at compile time~\cite{asan} 
can be added before any memory access to ensure 
validity of the memory going to be accessed. This can be accomplished via compiler-assisted program analysis. Traditionally, such memory-safe compilers will generate an exception and abort when such violations are found at run-time.

However, we make the observation that one could also react to such violations by {\em bypassing} the illegal instructions, \emph{i.e., instructions that attempt to access memory illegally,} and thus favoring availability of the system. This is the main intuition upon which our CIMA approach is based upon. In particular, CIMA is a resilient mitigation strategy that effectively and efficiently prevents memory-safety attacks and guarantees 
system availability with minimal overhead (8.06\%).



To realize the aforementioned intuition behind CIMA, we face several technical challenges. Firstly, it is infeasible (in general) to statically compute the exact set of illegal memory accesses in a program. Consequently, a fully static approach, which modifies the program to eliminate the illegal instructions from the program, is unlikely to be effective. Moreover, such an approach will inevitably face scalability bottlenecks due to its heavy reliance on sophisticated program analysis. Secondly, even if the illegal instructions are identified during execution, it is challenging to bypass the manifestation of illegal memory access. This is because, such a strategy demands full control to manipulate the normal flow of program execution. Finally, to bypass the execution of certain instructions, we need modifications to the control flow of the program. From a technical perspective, such modifications involve the manipulation of program control flow over multitudes of passes in mainstream compilers.  


To alleviate the technical challenges, CIMA systematically combines compile-time instrumentation and runtime monitoring to defend against memory-safety attacks. Specifically, at compile time, each instrumented memory access is guarded via a conditional check to detect its validity at runtime. In the event where the conditional check fails, CIMA skips the respective memory access at runtime. 
%
To implement such a twisted flow of control, CIMA automatically transforms the program control flow logic within mainstream compilers. This makes CIMA a proactive mitigation strategy against a large class of memory-safety attacks. 

It is the novel mitigation strategy that sets our CIMA approach apart from the existing works~\cite{asan,safedispatch,ropocop,cfi_cots,cfi_gcc,cfi_sp,softbound,cets,memsafe,drmemory,ccured,mudflap}. Most of the existing mitigation schemes against memory-safety attacks are primarily programmed to abort or restart the victim system when an attack or a memory-safety violation is detected. Other schemes, such as self-healing \cite{sting_song,vsef_song,vbs_song} or live patching \cite{live_patching1}, 
are based on directly detecting exploitations or attacks and resuming the corrupted system from a previous safe state. CIMA never aborts the system and avoids the heavy overhead of maintaining system states for checkpoints. On the contrary, CIMA follows a fundamentally different approach, i.e., to continue execution by skipping {\em only} the illegal instructions. 

Although our proposed security solution is applicable to any computing system that involves C/C++ programs, we mainly focus on the CPS domain. This is because, unlike the mainstream systems, CPS often imposes conflicting design constraints involving real-time guarantees, physical-state resiliency involving its physical dynamics and security. 
In CPS, the memory-safety vulnerabilities might be found in the firmware (or sometimes in the control software) of PLCs. Such firmware is commonly implemented in C/C++ for the sake of efficiency. Consequently, it is not uncommon to have buffer over/underflows and dangling pointers being regularly discovered even in modern PLCs. In fact, recent trends in Common Vulnerabilities and Exposures (CVEs) 
show the high volume of interest in exploiting these vulnerabilities in PLCs~\cite{cve_ab,cve_ab2,cve_ab3,cve_siemens1,cve_siemens2,cve_sem1,cve_sem2,cve_abb}. This shows that the mitigation of memory-safety attacks in CPS should not merely be restricted to academic research. Instead, it is a domain that requires urgent and practical security solutions to protect a variety of critical infrastructures at hand. Nonetheless, attacks that manipulate sensor and actuator values in CPS (either directly on sensor/actuator devices or on communication channels) are orthogonal issues that are out of our scope.

In summary, \emph{our work tackles the problem of ensuring critical systems and services to remain available and effective while successfully mitigating a wide-range of memory-safety attacks}.
%
We make the following technical contributions: \begin{enumerate}
\item We effectively prevent system crashes that could be arisen due to memory-safety violations.
\item We effectively and efficiently prevent memory-safety attacks in any computing system.
\item We define the notion of physical-state resiliency that is crucial for CPS and should be met alongside strong security guarantees. 
\item Our mitigation solution ensures physical-state resiliency and system availability with reasonable runtime and storage overheads. Thus, it is practically applicable to systems with stringent timing constraints, such as CPS. 
\item We evaluate the effectiveness and efficiency of our approach on two real-world CPS testbeds. 
\end{enumerate}

\section{Background}\label{sec:background}
In this section, we introduce the necessary background in the context of our CIMA approach. 

\subsection{CPS}\label{subsec:cps}
CPS constitutes of complex interactions between entities in the physical space and the cyber space over communication networks. Unlike traditional IT systems, such complex interactions are accomplished via communication with the physical world via sensors and with the digital world via controllers (PLCs) and other embedded devices.
CPS usually impose hard real-time constraints. 
If such real-time constraints are not met, then the underlying system might run into an unstable and unsafe state. Moreover, 
the devices in a typical CPS are also resource constrained. 
For example, PLCs and I/O devices have limited memory and computational power. In general, a typical CPS consists of the following entities: 



\begin{itemize}
\item {\bf Physical plants}: Physical systems where the actual processes take place.

\item {\bf Sensors}: Devices that are capable of reading or observing information from plants or physical processes. 

\item {\bf PLCs}: Controller devices that receive sensor inputs, make decisions and issue control commands to actuators.

\item {\bf Actuators}: Physical entities that are capable of implementing the control commands issued by the PLCs.

\item {\bf Communication networks}: The communication medium through which packets containing sensor inputs, control commands, diagnostic information and alarms transmit from one CPS entity to another.

\item {\bf SCADA}: A software entity designed for monitoring and controlling different processes in a CPS. It often comprises a human-machine interface (HMI) and a historian server. The HMI is used to display state information of plants and physical processes in the CPS. The historian server is used to store all operational data and the history of alarms.

\end{itemize}

An abstraction of a typical CPS architecture is shown in 
Figure~\ref{fig:cps_model}. In Figure~\ref{fig:cps_model}, $x$ denotes the physical state of the plant, $y$ captures the sensor measurements and $u$ denotes the control command computed by the PLC at any given point of time.


\subsection{ASan}\label{subsec:asan}%
Despite the presence of several memory error detector tools, their applicability in CPS is limited due to several reasons. 
This includes the lack of error coverage, significant performance overhead and other technical compatibility issues. After researching and experimenting on various memory-safety tools, we chose ASan~\cite{asan} as our memory error detector tool. 
Our choice is motivated by its broad error coverage, high detection accuracy and relatively low runtime overhead when compared with other code-instrumentation based tools~\cite{asan,asgithub}. 


ASan is a compile-time memory-safety tool based on code instrumentation. It instruments C/C++ programs at compile time. The instrumented program will then contain additional ASan libraries, which are checked to detect possible memory-safety violation at runtime. Such an instrumented code can detect buffer over/underflows, use-after-free errors (dangling pointers), use-after-return errors, initialization order bugs and memory leaks. 


Since ASan was primarily designed for x86 architectures, it has compatibility issues with RISC-based ARM or AVR based architectures. Therefore, we adapted ASan for ARM-based architecture in our system. 


\smallskip\noindent
\textbf{Limitations of ASan}:
Although ASan covers a wide range of memory errors, it does not cover some memory-safety errors such as uninitialized memory reads and some use-after-return bugs. In fact, such errors are less critical and rarely exploited in practice. ASan also has minor limitation in detection accuracy. Although it offers high detection accuracy for most memory-safety vulnerabilities, there are rare false negatives for global buffer overflow and use-after-free vulnerabilities. This might allow memory-safety attacks to bypass the checks enforced by ASan with low probability. The other major limitation of ASan is its ineffective mitigation strategy; it simply aborts the system whenever a memory-safety violation or an attack is detected. This makes ASan inapplicable in systems with stringent availability constraints. 

\smallskip\noindent
\textbf{ASan as a debugging and monitoring tool}:
Because of the limitations, as mentioned in the preceding paragraph,  ASan is often considered as rather a debugging tool than a runtime monitoring tool. However, using ASan only as a debugging tool does not guarantee memory-safety. Because, most memory-safety vulnerabilities 
(e.g. buffer overflows) can be probed by systematically crafted inputs by attackers who aim to exploit it. Since carefully tailored inputs might not be used during debugging, ASan might miss important memory bugs that can be exploited by an attacker at runtime. Therefore, only debugging the program does not offer sufficient guarantee in detecting critical memory-safety vulnerabilities.


In our CIMA approach, we adopt ASan for the dual purpose of debugging and runtime monitoring, with the specific focus on mitigating memory-safety vulnerabilities. As a runtime monitoring tool, CIMA leverages on ASan to detect attacker injected memory-safety bugs. Moreover, CIMA enhances the capability of ASan to mitigate memory-safety bugs on-the-fly. This is to ensure the availability of the underlying system and in stark contrast to plain system abort.  


\section{Attacker and system models}\label{sec:attacker_and_system_model}
In this section, we first discuss the attacker model. Then, we discuss the different traits of formally modeling our system and the related design constraints.  

\subsection{Attacker model}\label{attacker_model}

\begin{figure}[thb]
 \centering
 \includegraphics[scale=0.4]{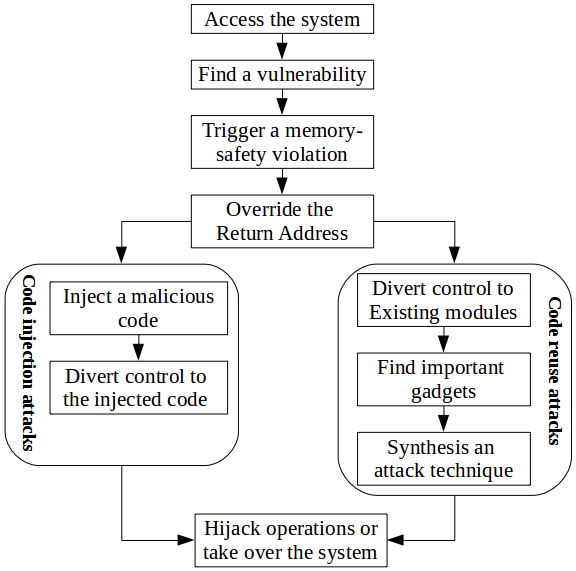}
 \caption{Overview of memory-safety attack exploitations}
 \label{fig:attacker_model}
\end{figure}

The main objective of memory-safety attacks, like code injection and code reuse, is to get privileged access or take control (otherwise to hijack/subvert specific operations) of the vulnerable system. To achieve this, vulnerabilities such as buffer overflows and dangling pointers of a program are targeted. For example, to exploit a buffer overflow, the attacker sends carefully crafted input to the buffer. When the buffer overflows, the attacker can manipulate important memory addresses, like return address of a function, and divert control flow of the program. A detailed account of such exploitations can be found in \cite{brop,coop}. 
In general, a typical memory-safety attack follows the following steps (See Figure \ref{fig:attacker_model}):
\begin{enumerate}
\item Interacting with the victim PLC, e.g., via network connection (for remote attacks). 
\item Finding a memory-safety vulnerability (e.g., buffer overflow, dangling pointers) in the PLC firmware or control software 
with the objective of exploiting it.
\item Triggering a memory-safety violation on the PLC, e.g., overflowing a buffer. 
\item Overwriting critical addresses of the vulnerable program, e.g., overwriting return address of the PLC program. 

\item Using the modified address, divert control flow of the program to an injected (malicious) code (i.e. code injection attacks) or to an existing module of the vulnerable program (i.e. code reuse attacks). In the former case, the attacker can take over the PLC with the injected malicious code. In the latter case, the attacker still needs to collect appropriate gadgets from the program (basically by scanning the program's text segment), then she will synthesize a shellcode that will allow her to take over the PLC.
\end{enumerate}


\subsection{Modeling CPS timing constraints}\label{modeling_cps_timing_constraints}
Most cyber-physical systems are highly time-critical. The communication between its different components, such as sensors, controllers (PLCs) and actuators, is synchronized by system time. Therefore, delay in these CPS nodes could result in disruption of the control system or damage the physical plant. 
In particular, PLCs form the main control devices and computing nodes of a typical CPS. As such, PLCs often impose hard real-time constraints to maintain the stability of the control system in a CPS. 
In the following section, we define and discuss the notion of real-time constraints imposed on a typical CPS. 


\paragraph*{\textbf{Modeling real-time constraints}}\label{modeling_real_time_constraints}

In general, PLCs undergo a cyclic process 
called \emph{scan cycle}. This involves three major operations: input scan, PLC program (logic) execution and output update. 
The time it takes to complete these operations is called \emph{scan time} ($T_{s}$).
A typical CPS defines and sets an upper-bound on time taken by PLC scan cycle, called \emph{cycle time} ($T_c$). This means the scan cycle must be completed within the duration of the cycle time specified, i.e., $T_{s} \leq T_c$. We refer this constraint as a \emph{real-time constraint} of the PLC. By design, PLCs meet this constraint. However, due to additional overheads, such as memory-safety overheads (MSO), PLCs might not satisfy its real-time constraint. As discussed above, by hardening the PLC with our memory-safety protection (detection + mitigation), the scan time increases. This increase in the scan time is attributed to the MSO. Concretely, the memory-safety overhead is computed as follows:
\begin{equation}\label{eq:mso}
\textit{MSO} = \hat{T}_{s} - T_{s},
\end{equation}
where $\hat{T}_s$ and $T_{s}$ are scan time with and without memory-safe compilation, respectively. A detailed account of modeling $T_{s}$ is provided in \REM{existing research work}our earlier works \cite{eyasu_cybericps,eyasu_essos}. 

It is crucial to check whether the induced MSO by the memory-safe compilation still satisfies the real-time constraint imposed by the PLC. To this end, we compute MSO for -- 1) average-case and 2) worst-case scenarios. In particular, after enabling memory-safe compilation, we compute the scan time (i.e. $\hat{T}_{s}$) for $n$ different measurements and compute the respective average and worst-case scan time. Formally, we say that the MSO is acceptable in average-case if the following condition is satisfied: 
\begin{equation}
\label{eq:average-case-tolerability}
\boxed{
\frac{\sum_{i = 1}^{n} \hat{T}_s(i)}{n} \leq T_c}
\end{equation}
In a similar fashion, MSO is acceptable in the worst-case with the following condition: 
\begin{equation}
\label{eq:worst-case-tolerability}
\boxed{
\displaystyle \max_{i = 1}^{n}\ \hat{T}_s(i) \leq T_c}
\end{equation}
where $\hat{T}_{s}(i)$ captures the scan time for the $i$-th measurement after the memory-safe compilation. 

\subsection{Physical-state resiliency}\label{modeling_physical_state_resiliency}

The stability of CPS controllers (i.e. PLCs in our case) plays a crucial role in enforcing the dynamics of a cyber-physical system to be compliant with its requirement. For example, assume that a PLC issues control commands every $T_c$ cycle and an actuator receives these commands at the same rate. Therefore, cycle time of the PLC is $T_c$. If the PLC is down for an arbitrary amount of time say $\tau$, then the actuator will not receive fresh control commands for the duration $\tau$. Consequently, the physical dynamics of the respective CPS will be affected for a total of $\frac{\tau}{T_c}$ scan cycles. 

We note that the duration $\tau$ might be arbitrarily large. Existing solutions~\cite{asan}, albeit in a non-CPS environment, typically revert to aborting the underlying process or restart the entire system when a memory-safety attack is detected. Due to the critical importance of availability constraints in CPS, our CIMA approach never aborts the underlying system. Nevertheless, CIMA induces 
an overhead to the scan time of the PLC, as discussed in the preceding section. Consequently, the scan time of PLCs, with CIMA-enabled memory-safety (i.e. $\hat{T}_s$), may increase beyond the cycle time (i.e. $T_c$). In general, to accurately formulate $\tau$ (i.e. the amount of downtime for a PLC), we need to consider the following three mutually exclusive scenarios: 
\begin{enumerate}
\item The system is aborted or restarted.
\item The system is neither aborted nor restarted and $\hat{T}_s \leq T_c$. In this case, there will be no observable impact on the physical dynamics of the system. This is because the PLCs, despite having increased scan time, still meet the real-time constraint $T_c$. Thus, they are not susceptible to downtime. 
\item The system is neither aborted nor restarted and $\hat{T}_s > T_c$. In this scenario, the PLCs will have a downtime of $\hat{T}_s - T_c$, as the scan time violates the real-time constraint $T_c$. 
\end {enumerate}

Based on the intuitions mentioned in the preceding paragraphs, we now formally define $\tau$, i.e., the downtime of a PLC as follows: 
\begin{equation}
\label{eq:downtime}
\tau = \begin{cases}
\Delta, &\text{system is aborted/restarted}\\
0, &\text{$\hat{T}_s \leq T_c$}\\
\hat{T}_s - T_c, &\text{$\hat{T}_s > T_c$}
\end{cases}
\end{equation}
where $\Delta$ captures a non-deterministic threshold on the downtime of PLCs when the underlying system is aborted or restarted.  

\paragraph*{\textbf{Example}}
As an example, let us consider the first process in SWaT (discussed in Section~\ref{subsec:swat}). This process controls the inflow of water from an external water supply to a raw water tank. PLC1 controls this process by opening (with ``ON'' command) and closing (with ``OFF'' command) a motorized valve, i.e., the actuator, connected with the inlet pipe to the tank. If the valve is ``ON'' for an arbitrarily long duration, then the raw water tank might overflow, often causing a severe damage to the system. This might occur due to the PLC1 downtime $\tau$, during which, the control command (i.e. ``ON'') computed by PLC1 may not change. Similarly, if the actuator receives the command ``OFF'' from PLC1 for an arbitrarily long duration, then the water tank may underflow. Such a phenomenon will still affect the system dynamics. This is because tanks from other processes may expect raw water from this underflow tank. 
In the context of SWaT, the {\em tolerability} of PLC1 downtime $\tau$ (cf. Eq.~(\ref{eq:downtime})) depends on the physical state of the water tank (i.e. water level) and the computed control commands (i.e. ON or OFF) by PLC1. In the next section, we will formally introduce this notion of tolerance, as termed \emph{physical-state resiliency}, for a typical CPS. 

\begin{figure}[t]
 \centering
 \includegraphics[scale=0.42]{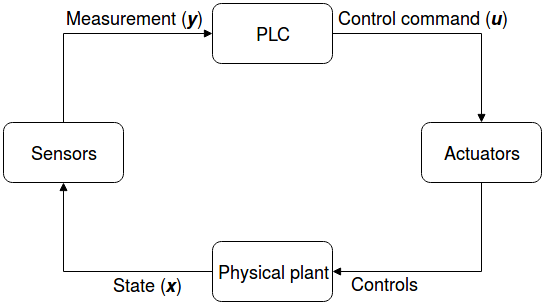}
 \caption{An abstraction of a CPS model}
 \label{fig:cps_model}
\end{figure}

\paragraph*{\textbf{Modeling physical-state resiliency}}
To formally model the physical-state resiliency, we will take a control-theoretic approach. For the sake of simplifying the presentation, we will assume that the dynamics of a typical CPS, without considering the noise and disturbance on the controller, is modeled via a linear-time invariant. This is formally captured as follows (cf. Figure~\ref{fig:cps_model}): 
\begin{equation}
\label{eq:modeling_state}
x_{t+1} = Ax_{t} + Bu_{t}
\end{equation}
\begin{equation}
\label{eq:modeling_output}
y_{t} = Cx_{t}
\end{equation}
where $t \in \mathbb{N}$ captures the index of discretized time domain. $x_t \in \mathbb{R}^{k}$ is the state vector of the physical plant at time $t$, $u_t \in \mathbb{R}^{m}$ is the control command vector at time $t$ and $y_t \in \mathbb{R}^{k}$ is the measured output vector from sensors at time $t$. $A \in \mathbb{R}^{k \times k}$ is the state matrix, $B \in \mathbb{R}^{k \times m}$ is the control matrix and $C \in \mathbb{R}^{k \times k}$ is the output matrix. 



We now consider a duration $\tau \in \mathbb{R}$ for the PLC downtime. To check the tolerance of $\tau$, we need to validate the physical state vector $x_t$ at any discretized time index $t$. To this end, we first assume an upper-bound $\omega \in \mathbb{R}^{k}$ on the physical state vector $x_t$ at an arbitrary time $t$. Therefore, for satisfying physical state resiliency, $x_t$ must not exceed the threshold $\omega$. In a similar fashion, we define a lower-bound $\theta \in \mathbb{R}^{k}$ on the physical state vector $x_t$. 

With the PLC downtime $\tau$, we revisit Eq.~(\ref{eq:modeling_state}) and the state estimation is refined as follows: 
\begin{equation}
\label{eq:modeling_state_in_delay}
x'_{t+1} = Ax_{t} + Bu_{t-1}[\![ t, t+\tau ]\!]
\end{equation}
where $x'_{t+1} \in \mathbb{R}^{k}$ is the estimated state vector at time $t+1$ and the PLCs were down for a maximum duration $\tau$. The notation $u_{t-1}[\![ t, t+\tau ]\!]$ captures that the control command $u_{t-1}$ was active for a time interval $[t, t+\tau]$ due to the PLC downtime. In Eq.~(\ref{eq:modeling_state_in_delay}), we assume, without loss of generality, that $u_{t-1}$ is the last control command received from the PLC before its downtime. 

Given the foundation introduced in the preceding paragraphs, we say that a typical CPS (cf. Figure~\ref{fig:cps_model}) satisfies physical-state resiliency if and only if the following condition holds at an arbitrary time index $t$: 
\begin{equation*}
\theta \leq  x'_{t+1} \leq \omega
\end{equation*}
\begin{equation}
\boxed{
\label{eq:cls1}
\theta \leq Ax_{t} + Bu_{t-1}[\![ t, t+\tau ]\!] \leq \omega}
\end{equation}

\begin{figure}[h]
 \centering
 \includegraphics[scale=0.3]{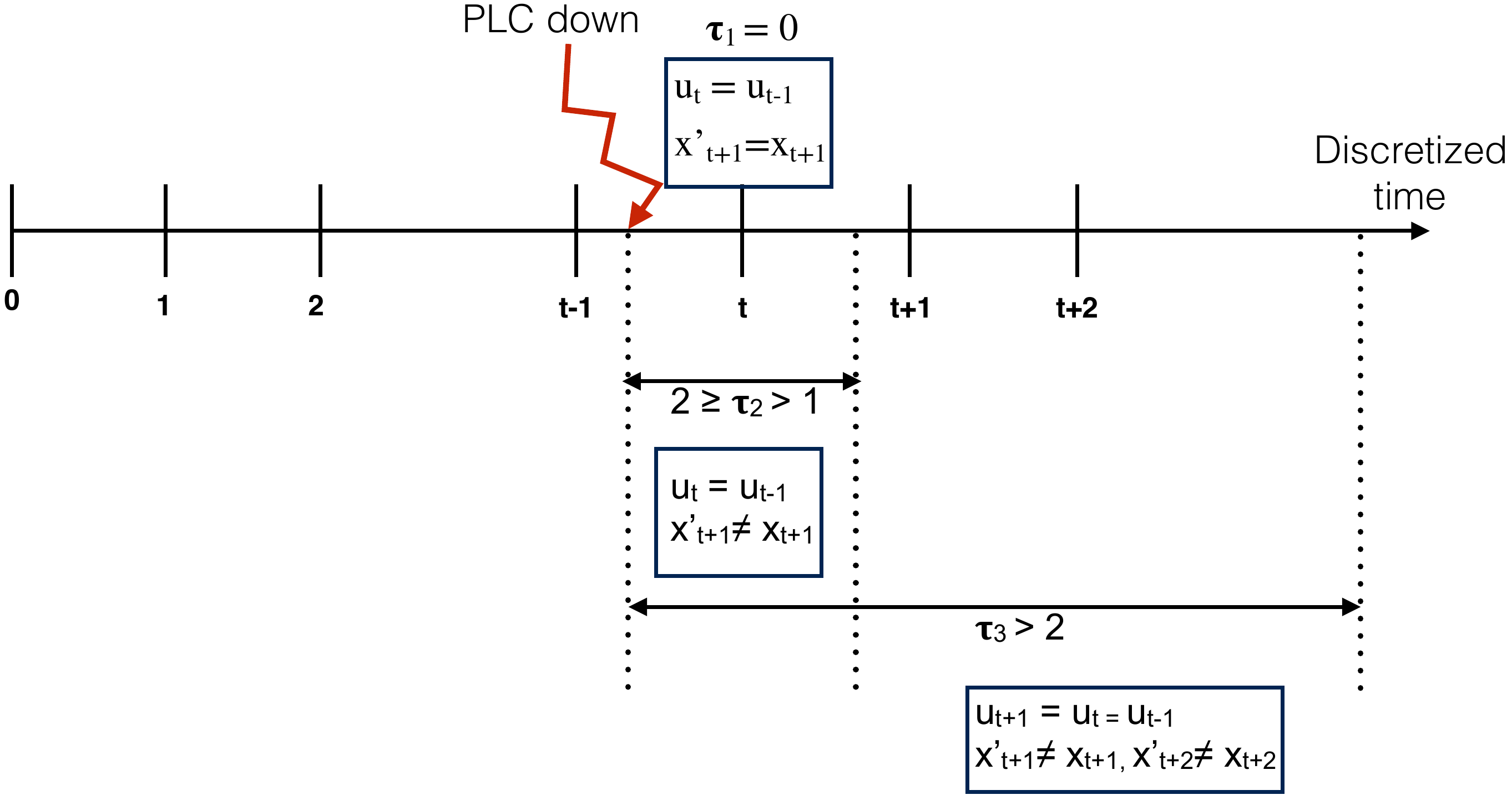}
 \caption{Illustrating the impact of PLC downtime}
 \label{fig:downtime}
\end{figure}

Figure~\ref{fig:downtime} illustrates three representative scenarios to show the consequence of Eq.~(\ref{eq:cls1}). If the downtime $\tau_1 =0$, then $u_t$ (i.e. control command at time $t$) is correctly computed and $x'_{t+1}=x_{t+1}$. If the downtime $\tau_2 \in (1,2]$, then the control command $u_{t}$ will be the same as $u_{t-1}$. Consequently, $x'_{t+1}$ is unlikely to be equal to $x_{t+1}$. Finally, when downtime $\tau_3 > 2$, the control command vector $u_{t+i}$ for $i \ge 0$ will be the same as $u_{t-1}$. As a result, the estimated state vectors $x'_{t+j}$ for $j \ge 1$ will unlikely to be identical to $x_{t+j}$. 




\section{{\bf CIMA}: {\bf C}ountering {\bf I}llegal {\bf M}emory {\bf A}ccesses}\label{sec:cima}
CIMA is a mitigation technique designed to counter illegal memory accesses at runtime. We first outline a high-level overview and the key ingredients of our approach. Later in this section, we discuss the specific implementation traits in more detail. 

\subsection{Overview of CIMA}
\label{cima_approach}

\subsubsection*{\textbf{Objective}}
CIMA follows a proactive approach to mitigate memory-safety attacks and thereby ensuring system availability and physical-state resilience in CPS. The key insight behind CIMA is to prevent any operation that attempts to illegally access memory. 
We accomplish this by skipping (i.e. not executing) the illegal instructions 
that attempt to access memory illegally. 
For example, consider the exploitation of a memory-safety attack shown in Figure~\ref{fig:attacker_model}. With our CIMA approach, the attack will be ceased at step 3 (i.e. ``Trigger a memory-safety violation") of the exploitation process. 

\subsubsection*{\textbf{Workflow of CIMA}}
CIMA systematically manipulates the compiler-generated control flow graph (CFG) to accomplish its objective, i.e., to skip illegal memory accesses at runtime. 
Control flow graph (CFG) of a program is represented by a set of basic blocks having incoming and outgoing edges. A basic block~\cite{basic_block} is a straight-line sequence of code with only one entry point and only one exit, but it can have multiple predecessors and successors. 
CIMA works on a common intermediate representation of the underlying code, thus making our approach applicable for multiple high-level programming languages and low-level target architectures. 

\begin{figure}[thb]
 \centering
 \includegraphics[scale=0.34]{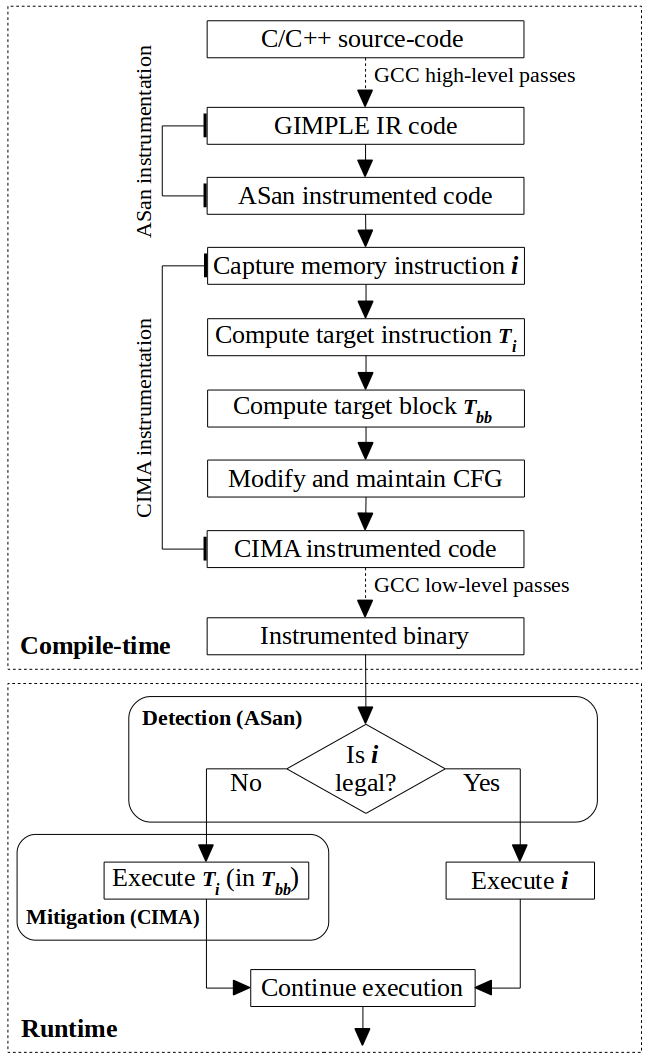}
 \caption{A high-level architecture of CIMA}
 \label{fig:cima_architecture}
\end{figure}

To manipulate the CFG, CIMA needs to instrument the set of memory-access operations in such a fashion that they trigger memory-safety violations at runtime. To this end, CIMA leverages off-the-shelf technologies, namely address sanitizers (ASan). A high-level workflow of our entire approach appears in Figure~\ref{fig:cima_architecture}. 
The implementation of CIMA replaces the ineffective mitigation strategy (i.e. system abort) of ASan  with an effective, yet lightweight scheme. Such a scheme allows the program control flow to jump to the immediately next instruction that does not exhibit illegal memory access. In such a fashion, CIMA not only ensures high accuracy of memory-safety attack mitigation, but also provides confidence in system availability and resilience. 
In the following section, we describe our CIMA approach in detail.  

\subsection{Approach of CIMA}
\label{sec:methodology}

An outline of our CIMA approach appears in Algorithm~\ref{cima:transformation}. As input, CIMA takes a function (in the GIMPLE format) instrumented by ASan. As output, CIMA generates a GIMPLE function with memory-safety mitigation enabled. We note that the modification is directly injected in the compiler workflow. To this end, CIMA modifies the internal data structures of GCC to reflect the changes of control flow. 
Specifically, CIMA manipulates the intermediate representation, which is often derived in the static-single-assignment (SSA) form. To this end, CIMA also ensures that the SSA form is maintained during the manipulation, so as not to disrupt the compiler workflow. We now elaborate the crucial steps of CIMA in detail and discuss the specific choices taken during its implementation.

\begin{algorithm}[thb]
\small
\SetKwProg{Fn}{}{ \{}{\}}
\KwIn{Function $\textit{fun}()$ with ASan Instrumentation}
\KwOut{Function $\textit{funCIMA}()$ with CIMA-enabled memory safety}
 \ForAll { basic block $\mathbf{bb}$ in $\textit{fun}()$} {
 \ForAll { instruction $\mathbf{i}$ instrumented by ASan in $\mathbf{bb}$} { \label{cima:iteratebb}
 Let $\mathit{check}_{bb}$ holds the 
 memory-safety check condition $\mathit{check}_{i}$ for instruction $\mathbf{i}$ in $\mathbf{bb}$\\
 Let $\mathit{abort}_{bb}$ be the basic block where control reaches to when instruction $\mathbf{i}$ is illegal\\
 Find target instruction $\mathbf{{T}_i}$ for instruction $\mathbf{i}$\\
  \eIf{$\mathbf{i}$ and $\mathbf{{T}_i}$ reside in the same basic block $\mathbf{bb}$} {
    $temp_{bb}$ := $succ(bb)$\\
    \textsf{\bf /* split basic block */}\\ 
 	Split $\mathbf{bb}$ into basic blocks $i_{bb}$ and $T_{bb}$\\ \label{bb_spliting_1}
	$i_{bb}$ holds instructions of $\mathbf{bb}$ up to $\mathbf{i}$\\  \label{bb_spliting_2}
    $T_{bb}$ holds instructions from $\mathbf{{T}_i}$ up to the end of $\mathbf{bb}$\\  \label{bb_spliting_3}
    \textsf{\bf /* Modify control flow */} \\
    \textsf{/* $succ(bb)$ captures successor of $\mathbf{bb}$ */}\\ 
    $succ(i_{bb})$ := $\{T_{bb}\}$\\ \label{modify_cfg_1}
    $succ(check_{bb})$ := $\{i_{bb}, T_{bb}\}$\\ 
    $succ(T_{bb})$ := $temp_{bb}$\\ \label{modify_cfg_2}
}
{
    \textsf{\bf /* Modify control flow */}\\ 
	Let $\mathbf{{T}_i}$ be in basic block $T_{bb}$\\
    $succ(check_{bb})$ := $\left ( succ(check_{bb}) \setminus \{abort_{bb}\} 
    \right ) \cup \{T_{bb}\}$ \label{modify_cfg_3}\\
}

		}
    }
 \caption{CIMA's approach to ensure memory safety}
 \label{cima:transformation}
 \end{algorithm}



\smallskip\noindent
\textbf{Capture memory access instructions}:
CIMA validates memory accesses at runtime to detect and mitigate illegal memory access instructions. To achieve this, CIMA combines a compile-time code instrumentation and a runtime validity check technique. The code instrumentation includes instrumenting memory addresses (via ASan) and memory access instructions (via CIMA). 
The memory address instrumentation creates poisoned (i.e. illegal) memory regions, known as {\em redzones}, around stack, heap and global variables. Since these redzones are inaccessible by the running program, any memory instruction, attempting to access them at runtime will be detected as an illegal instruction. 

To mitigate the potential illegal instructions, CIMA instruments memory access instructions at compile-time. To this end, CIMA captures memory access instructions from the ASan instrumented code. These instructions serve as the potential candidates for illegal instructions at runtime. 



\smallskip\noindent
\emph{\bf Compute target instruction}: 
To prevent the execution of an illegal memory access instruction $i$, CIMA computes its corresponding target instruction $T_{i}$. We note that the set of potentially illegal instructions are computed via the technique as explained in the preceding paragraph. {\em The target instruction $T_i$ for an illegal instruction $i$ is the instruction that will be executed right after $i$ is bypassed}. 
The compile-time analysis of CIMA generates an instrumented binary in such a fashion that if an illegal instruction $i$ is detected at runtime, then $i$ is bypassed and control reaches target instruction $T_i$.
If $T_{i}$ is detected as illegal too, then the successor of $T_{i}$ will be executed. This process continues until an instruction is found without illegal memory access at runtime. 

\begin{figure*}[t]
\begin{center}
\resizebox{\textwidth}{!}{%
\begin{tabular}{cccc}
\rotatebox{0}{
\includegraphics[scale = 0.23]{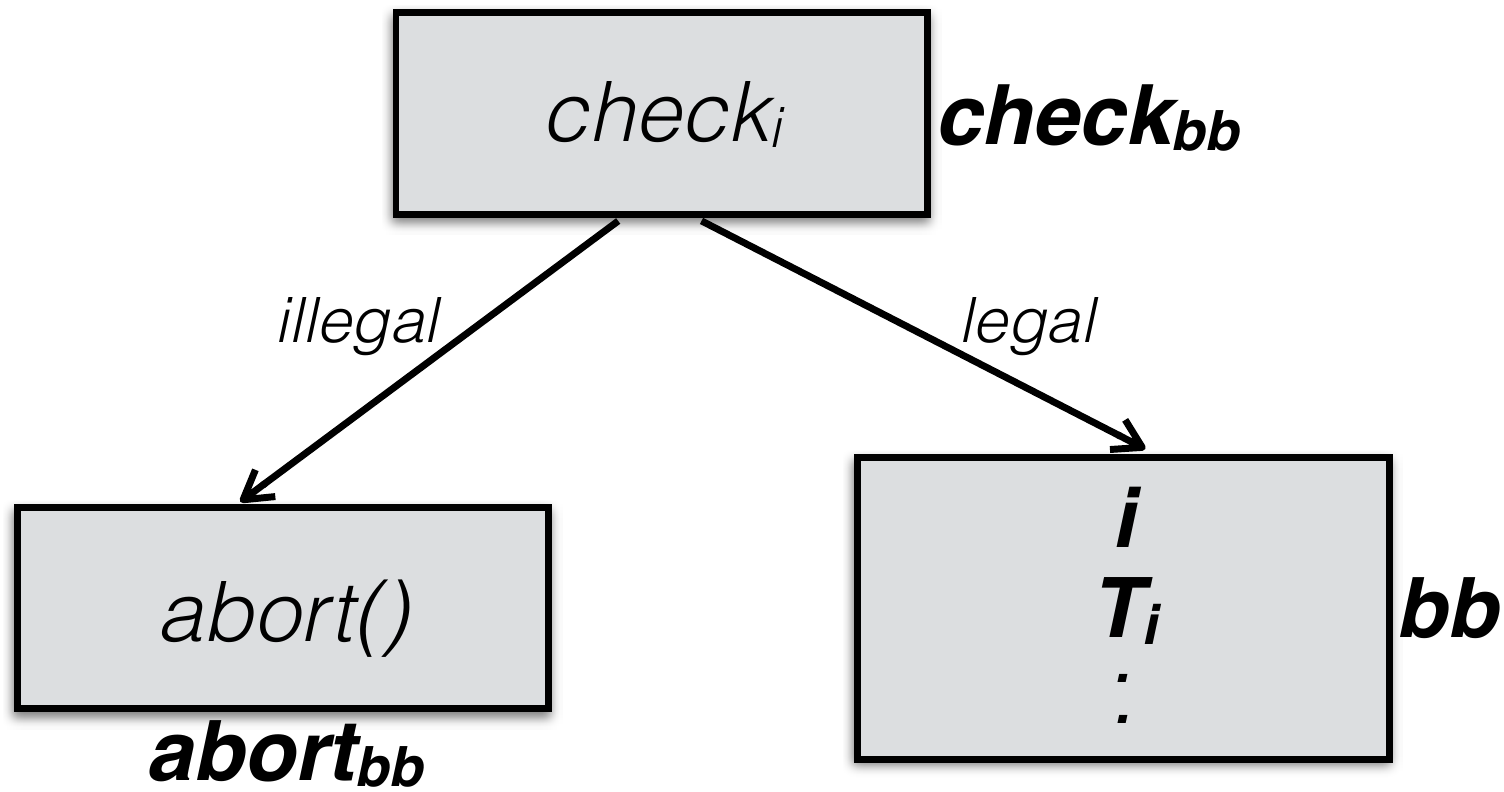}} &
\rotatebox{0}{
\includegraphics[scale = 0.23]{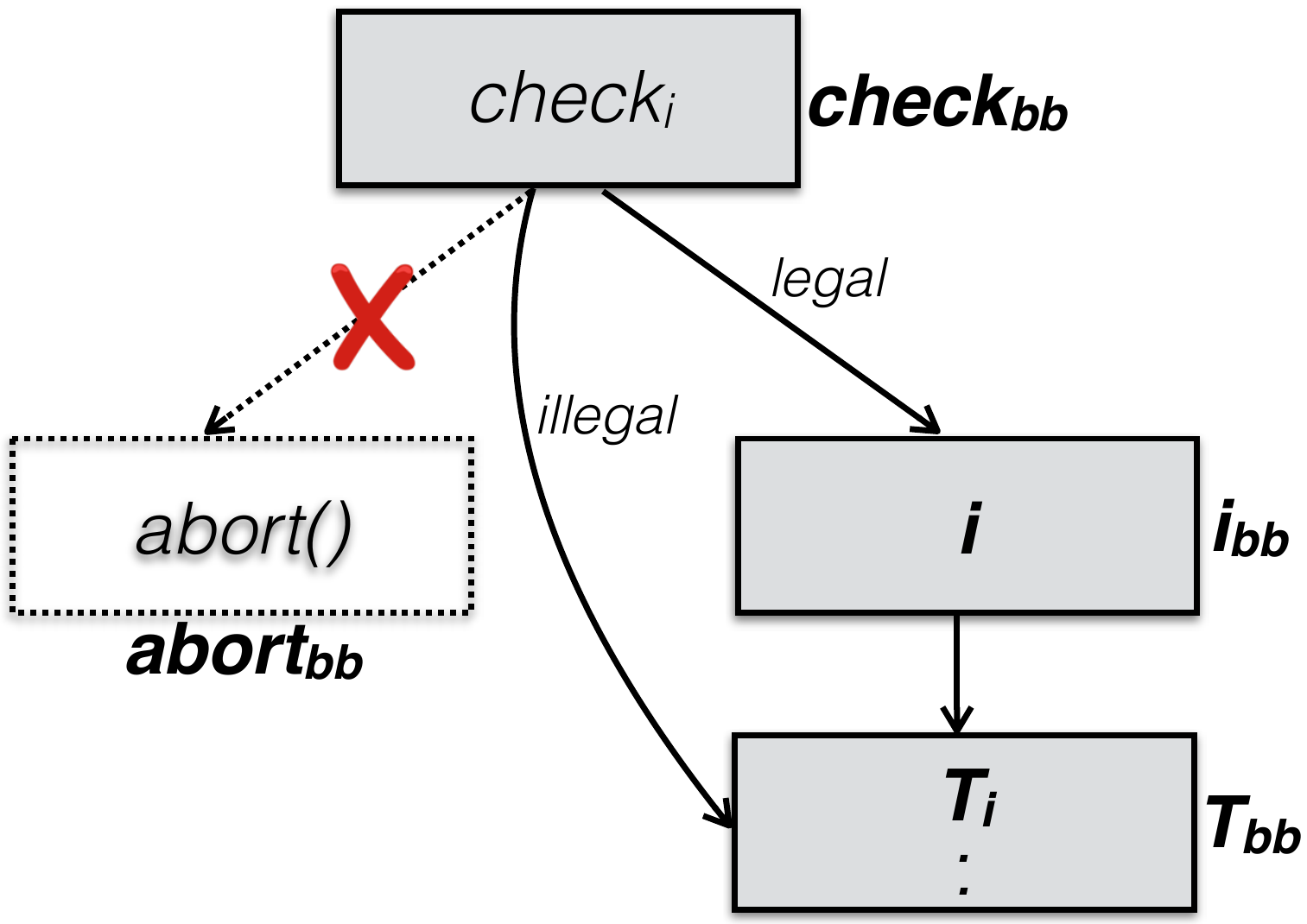}} & 
\rotatebox{0}{
\includegraphics[scale = 0.23]{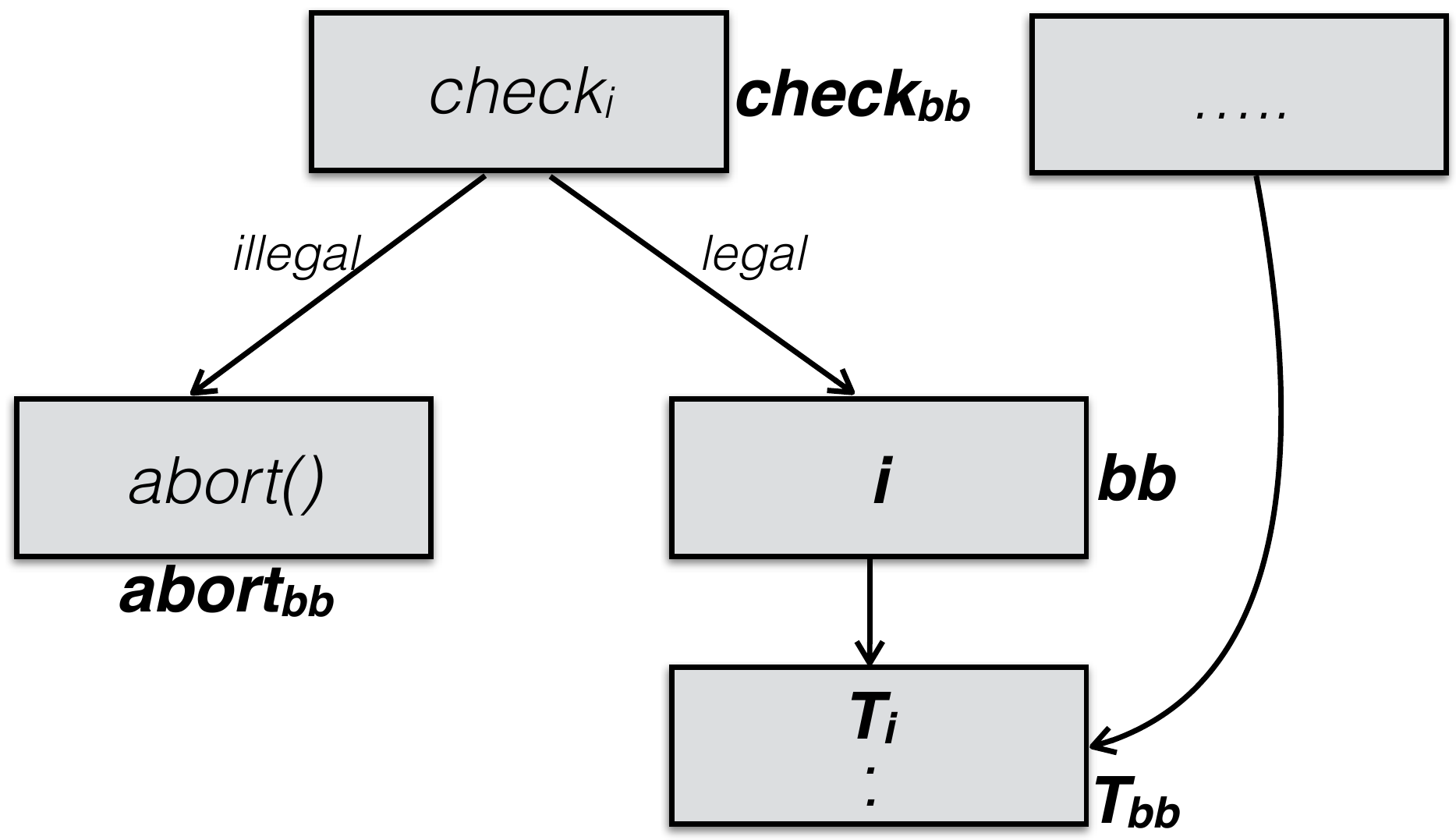}} & 
\rotatebox{0}{
\includegraphics[scale = 0.23]{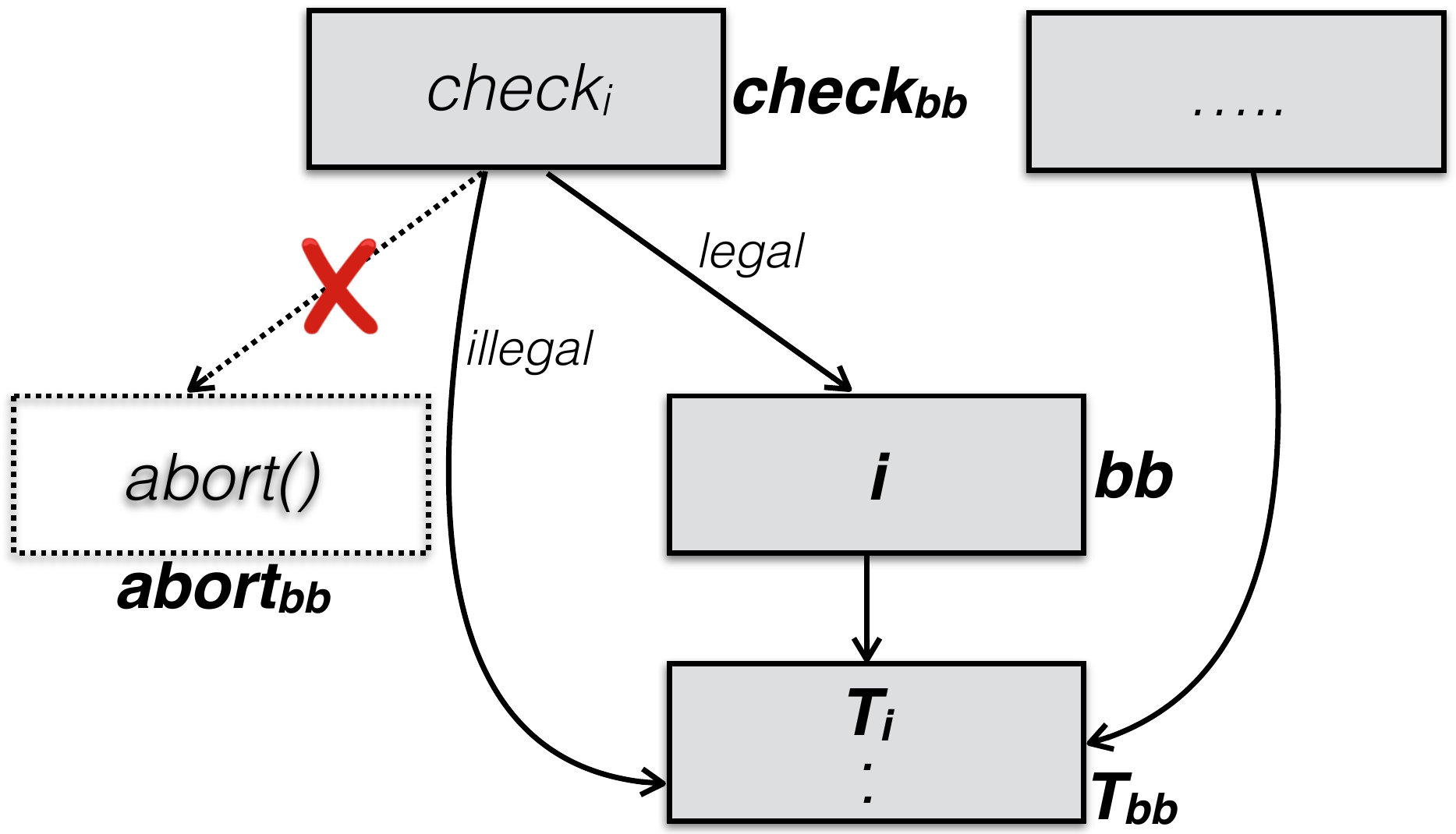}}\\
\textbf{(a)} & \textbf{(b)} & \textbf{(c)} & \textbf{(d)}
\end{tabular}
}
\end{center}
\vspace*{-0.1in}
\caption{Different scenarios encountered in CIMA's approach. $check_i$ is the inserted conditional check by ASan to identify illegal memory accesses in 
instruction $i$. 
$abort_{bb}$ is the basic block to which control jumps to if 
$i$ exhibits illegal memory access. {\bf (a):} ASan instrumented CFG where both 
$i$ and $\mathit{T}_i$ belong to the same basic block; 
{\bf (b):} the modification of control flow by CIMA for the ASan instrumented code in (a); 
{\bf (c):} ASan instrumented CFG where 
$i$ and $\mathit{T}_i$ belong to different basic blocks; 
{\bf (d):} the modification of control flow by CIMA for the ASan instrumented code in (c).
}
\label{fig:cima-example}
\vspace{-0.3cm}
\end{figure*}

Computing the target instruction is a critical step in our CIMA approach. 
The target instruction $T_{i}$ is computed as the successor of the 
memory access instruction $i$ in the CFG. 
We note that a potentially illegal instruction must access memory, as the objective of CIMA is to mitigate memory-safety attacks. 
At the GIMPLE IR-level, any memory access instruction 
has a single successor. Such a successor can either be the next instruction in the same basic block (see Figure~\ref{fig:cima-example}(a)) or the first instruction of the successor basic block (see Figure~\ref{fig:cima-example}(c)). Since CIMA works at the GIMPLE IR-level, it can identify the target instruction $T_i$ for any 
instruction $i$ by walking through the static control flow graph. 

We modify the control flow graph in such a fashion that the execution is diverted to $T_{i}$ (i.e. the successor of $i$) at runtime when instruction $i$ is detected to be illegal at runtime. However, the modification of the CFG depends on the location of the illegal and target instructions as discussed below. 




\smallskip\noindent
\emph{\bf Compute target basic block}: 
From the discussion in the preceding paragraph, we note that the execution of 
memory access instruction $i$ is conditional (depending on whether it is detected as illegal at runtime). In the case where the instruction exhibits an illegal memory access, a jump to the respective target instruction $T_{i}$ is carried out. However, it is not possible to simply divert the execution flow to the target instruction $T_{i}$. This needs to be accomplished via a systematic modification of the original program CFG at compile time. 

Consider the case when the illegal instruction $i$ and the target instruction $T_{i}$ reside in the same basic block (say $bb$). In this case, since the execution of $i$ is conditional, it is always the first instruction of its holding basic block whereas the target instruction $T_{i}$ appears as the next instruction in the basic block $bb$ (see Figure~\ref{fig:cima-example}(a)). However, it is not possible to make a conditional jump to $T_{i}$ within the same basic block $bb$, as this breaks the structure of the control flow graph. 
%
Thus, to be able to make a conditional jump to $T_{i}$, a target basic block (say $T_{bb}$), that contains $T_{i}$ as its head instruction, is created. 
In particular, we split the basic block $bb$ into two basic blocks -- $i_{bb}$ and $T_{bb}$ (cf. Algorithm \ref{cima:transformation}, Lines \ref{bb_spliting_1} -- \ref{bb_spliting_3}). $i_{bb}$ holds the potentially illegal instruction $i$ and $T_{bb}$ holds the instructions of $bb$ following $i$, i.e., starting from the target instruction $T_{i}$ and up to the last instruction of $bb$. As such, 
control jumps to $T_{bb}$ if instruction $i$ is detected as illegal at runtime. 

If the illegal instruction $i$ and the respective target instruction $T_{i}$ are located in different basic blocks in the original CFG (see Figure~\ref{fig:cima-example}(c)), then 
there is no need to split any basic block. In such a case, the target basic block $T_{bb}$ will be the basic block holding $T_{i}$ as its first instruction. Therefore, CIMA diverts the control flow to $T_{bb}$, should $i$ exhibits an illegal memory access at runtime. 



 

\smallskip\noindent
\emph{\bf Modify and maintain CFG}: CIMA ensures the diversion of control flow of the program whenever an illegal memory access is detected. To accomplish such a twisted control flow, CIMA directly modifies and maintains the CFG (cf. Algorithm~\ref{cima:transformation}, Lines (\ref{modify_cfg_1} -- \ref{modify_cfg_2}) and (\ref{modify_cfg_3})), as explained in the preceding paragraph. Figure~\ref{fig:cima-example} illustrates excerpts of control flow graphs that are relevant to the modification performed by CIMA. In particular, Figure~\ref{fig:cima-example} demonstrates how the control flow is systematically manipulated to guarantee the system availability, while still mitigating memory-safety attacks.

\subsection{Illustrative Example}\label{sec:example-cima}
CIMA detected and mitigated two global buffer overflow vulnerabilities on the firmware of OpenPLC controller. Due to  space limitation, we discuss here only one such vulnerability. \REM{As shown in Program~\ref{openplc_vuln},}
A code fragment relevant to the vulnerability is shown in Program~\ref{openplc_vuln}. In Line~\ref{openplcvul:arrdef}, 
a buffer ``{\it int\_memory[]}'' (with a buffer size of 1024) is declared in the ``{\it glue\_generator.cpp}'' file. This buffer is also used in the ``{\it modbus.cpp}'' file, enclosed within a {\it for} loop (Lines~\ref{openplcvul:forbegin} -- \ref{openplcvul:forend}) under the ``{\it mapUnusedIO()}'' function. In the loop, the memory write operation ``{\it int\_memory[i] =} \&{\it mb\_holding\_regs[i]}'' (Line~\ref{openplcvul:forerror}) writes data to the buffer. However, due to a coding error, this operation exhibits a memory-safety violation. Such a violation occurs in the $1024^{th}$ iteration when the operation attempts to write data beyond the buffer limit. 
CIMA successfully mitigates such memory-safety violation. This was possible as the illegal memory access operation was bypassed in each iteration starting from the $1024^{th}$ iteration of the loop.

\begin{cppalgorithm}
\small 

//------------{\bf //glue\_generator.cpp}--------------------------\\
\#define BUFFER\_SIZE		1024\\
IEC\_UINT *int\_memory[BUFFER\_SIZE]; \label{openplcvul:arrdef}\\
IEC\_UINT *int\_output[BUFFER\_SIZE];\\

.\\
.\\

//------------{\bf //modbus.cpp}------------------------------------\\
\#define MIN\_16B\_RANGE			1024\\
\#define MAX\_16B\_RANGE			2047\\

\#define MAX\_HOLD\_REGS 			8192\\

IEC\_UINT mb\_holding\_regs[MAX\_HOLD\_REGS];\\
.\\
.\\
\textsf{/* This function sets the internal OpenPLC buffers to */}\\ 
\textsf {/* point to valid positions on the Modbus buffer */}

\SetKwProg{Fn}{}{ \{}{\}}

\Fn(){void mapUnusedIO()}{
    
    
    .\\
    .\\
	\For {int i = 0; i \textless= MAX\_16B\_RANGE; i++}{ \label{openplcvul:forbegin}
		\If {i \textless MIN\_16B\_RANGE}{
			\If {int\_output[i] == NULL}{
            	int\_output[i] = \&mb\_holding\_regs[i];
            }
		}
		\If {i \textgreater= MIN\_16B\_RANGE \&\& i \textless= MAX\_16B\_RANGE}{
			\If {int\_memory[i - MIN\_16B\_RANGE] == NULL}{ 
            	{\bf int\_memory[i] = \&mb\_holding\_regs[i]};\label{openplcvul:forerror}
			}
        }
    }\label{openplcvul:forend}
}
\caption{The OpenPLC vulnerability}
\label{openplc_vuln}
\end{cppalgorithm}

\section{Validating CIMA}
\label{sec:cima-validation}


In this section, we discuss the impact on using CIMA to mitigate memory-safety attacks. 

\subsubsection{\emph{\bf Breaking program semantics}}\label{breaking_program_semantics}

The mitigation strategy of CIMA by itself is based upon breaking semantics of the program when an invalid memory access is detected. This is accomplished via skipping the respective memory access. In the presence of memory-safety attacks, we believe it is extremely time and memory 
consuming (given the CPS context) to roll back to a semantics preserving state. 
Thus, CIMA aims for a lightweight solution (attributes to only 8\% overhead in real-world CPS) that works for most common cases and we validate CIMA with a real-world CPS.
Nevertheless, there exists semantics-preserving issues that deserve discussion. 
%
%
For example, in Program~\ref{cima_corner_cases}, Line \ref{variable_assignemnt}, variable {\it y} is dependent on the output of memory access instruction {\it arr[i]}. 
CIMA always checks the validity of the memory access {\it arr[i]} whenever the access occurs. If the access {\it arr[i]} is found to be illegal, CIMA skips it and the variable assignment, i.e., {\it y = arr[i]}, is also skipped as a side-effect. In this case, {\it y} preserves the last ``legally'' assigned value to it. Variable $y$ preserves its initial value if the access {\it arr[i]} is detected as illegal at the beginning (i.e. at index $i = 0$) otherwise. 
Concretely, the assignment of {\it y} can be recursively defined as follows:
\begin{equation}
\label{eq:io-dependency}
y = \begin{cases}
arr[i], &\text{if $arr[i]$ is legal instruction}\\
arr[i-1], &\text{if $arr[i]$ is illegal, $i > 0$}\\
y_{0}, &\text{if $arr[0]$ is illegal}
\end{cases}
\end{equation}
where $y_{0}$ is the initial value of {\it y}.

At a broader level, CIMA may encounter memory safety attacks either due to an inherent vulnerability in the program or due to an illegal memory access caused by the attacker (e.g. via well crafted attack inputs). For inherent program vulnerabilities, we hypothesize that the underlying program semantics was already flawed and therefore, skipping the illegal memory accesses (due to such vulnerabilities), despite affecting the program semantics, will not further impair the system functionality. For illegal memory accesses via carefully crafted attack inputs, we hypothesize that in most common cases the program functionality will be unaffected when CIMA skips such illegal memory accesses. This is due to the reason that such illegal memory accesses were executed for the sole purpose of launching memory-safety attacks and skipping them should not affect the functionality of the system.

\subsubsection{{\bf Continuity of program execution}}
\label{continuity_of_program_execution}
CIMA might not ensure continuity in program execution when a {\it loop bound} or {\it loop counter} is controlled by an attacker. Let us consider the example shown in Program~\ref{cima_corner_cases}, Line \ref{loop_bound}, where the loop bound {\it arr[x]} is dependent on the attacker input {\it x}. 
We note that the value of {\it x} determines whether the memory read instruction {\it arr[x]} is legal or not. CIMA skips the access to {\it arr[x]} should it turns out to be illegal. This could lead to premature abortion of the loop (e.g. if {\it arr[x]} is detected as illegal before the loop starts).

Consider a different scenario when the the loop counter $i$ is incremented by {\it arr[z]} (cf. Program~\ref{cima_corner_cases}, Line \ref{counter_incrementation}). Assuming {\it z} can be controlled by the attacker, {\it arr[z]} might manifest illegal memory access. Thus, CIMA bypasses the respective memory access and the value of {\it i}  remains unchanged. This leads to an execution that never terminates. 
\begin{cppalgorithm}
\small 
\SetKwProg{Fn}{}{ \{}{\}}
\Fn(){main()}{
	int arr[100], i, x, y, z;\\
    printf(``Enter the value of x and z:");\\
  	scanf(``\%d \%d", \&x, \&z);\\
	\For {i=0; i \textless {\bf arr[x]}; }{ \label{loop_bound}
		// Do something\\
        y = arr[i];\\ \label{variable_assignemnt}
        {\bf i += arr[z]} //incrementing the loop counter \label{counter_incrementation} 
	}
    return 0;
}
\caption{Attacker controlled loop bound and counter}
\label{cima_corner_cases}
\end{cppalgorithm}
%
%

The examples discussed in the preceding paragraphs may not happen in practice. This is because providing {\it loop bound} and {\it loop counter} values via attacker controlled memory accesses (e.g. {\it arr[x]} in Program~\ref{cima_corner_cases}) is certainly considered as a {\it bad coding practice}. 
The occurrence of such corner cases could result in an undesirable outcome (e.g. program crash) even in the absence of CIMA. 
In essence, avoiding such bad coding practices will also help 
CIMA to maintain continuity 
of program execution in the presence of memory-safety violations or attacks. 


\subsubsection{\emph{\bf Affecting the system dynamics}}\label{affecting_system_dynamics}

CIMA may affect the overall system dynamics (or the physical-state resiliency in the context of CPS) when it skips too many illegal memory access instructions. This is because skipping too many illegal instructions creates a delay in program execution. This, in turn, may result in the following scenarios:
\begin{itemize}
\item {\it Lack of fresh input}: Skipping instructions results impeding of fresh input to the system. For example, actuators in CPS may not receive a new control command while instructions are skipped due to memory-safety violation. This, in turn, may affect the system dynamics. 

\item {\it Denial-of-service (DoS) attack}: When CIMA skips too many illegal instructions, the program execution may get stuck for a long period. This may result in the denial-of-service.  
\end{itemize}

As an example, consider the case where CIMA detected the OpenPLC vulnerability (cf. Program~\ref{openplc_vuln}, Line \ref{openplcvul:forerror}). CIMA skipped the illegal memory write instruction (i.e. {\it int\_memory[i] = \&mb\_holding\_regs[i]}) 1024 times in the {\it for} loop. This creates a delay equivalent to 1024 iterations of the loop. 
%
Let us assume $\delta$ captures the total elapsed time to skip illegal instructions in the loop. During the period of $\delta$, the PLC is busy skipping invalid memory accesses and does not issue a control command to the actuator. The number of PLC scan cycles (similarly the number of control commands) missed during this delay can be computed as $\delta/T_{c}$. As such, the effect of $\delta$ in the CPS dynamics (or the physical-state resiliency) is analogous to the downtime (i.e. $\tau$) of the PLC discussed on Section~\ref{modeling_physical_state_resiliency}. Therefore, quantifying the level of delay ($\delta$) that is tolerable to satisfy physical-state resiliency of a typical CPS can be similarly modeled using Eq.~(\ref{eq:cls1}) as follows: 
\begin{equation}
\boxed{
\label{eq:cima_delay}
\theta \leq Ax_{t} + Bu_{t-1}[\![ t, t+\delta ]\!] \leq \omega}
\end{equation}


To minimize the effect of CIMA on system dynamics (i.e. to satisfy Eq.~(\ref{eq:cima_delay})), we propose the following two approaches:
\begin{itemize}
\item \emph{\bf Detect long skips with debugging}: As discussed in  Section~\ref{subsec:asan}, we make use of CIMA 
both as a debugging and a runtime memory-safety tool. If the illegal memory access occurs due to an existing vulnerability in a loop, i.e., the vulnerability is from the existing source-code of the program and not attacker injected, then CIMA automatically detects this vulnerability while debugging the program. 
This is experimentally validated by accurately detecting the vulnerabilities found on the OpenPLC firmware. We also developed an informative report (supported by an alarm) for detected and mitigated illegal memory access instructions. Therefore, the inherent vulnerabilities in the source-code should be manually fixed once they are discovered by our framework. On the other hand, as discussed on Section~\ref{continuity_of_program_execution}, attackers may manipulate memory accesses in the loop bound or loop counter via untrusted inputs. However, 
as discussed in the preceding sections, such vulnerabilities are caused by {\it bad coding practices} and they can be solved by avoiding bad coding practices. 

For illegal memory accesses that occur outside loops (e.g. a substantial number of illegal memory access instructions in the source-code), the skipping time is unlikely to affect the system dynamics. As evidenced by our experiments, the skipping time of a single instruction is negligible. As such, skipping thousands of instructions is still tolerable in the context of real-world CPS. 
Yet, it is unlikely to have tens of thousands of illegal memory accesses in the absence of loops.


\item \emph{\bf Exiting loop}: In certain cases, it might be possible to exit the loop when a sufficient number of illegal memory accesses are detected inside it. This will reduce the delay $\delta$.
However, more involved analysis are required to identify potential loops that can be skipped altogether as soon as a certain number of illegal memory accesses are detected within it. We plan to extend CIMA along this direction in the future. 
\end{itemize}

\section{case Studies and Experimental Design}\label{sec:experimental_design} 
\subsection{SWaT}\label{subsec:swat} 

SWaT~\cite{swat} is a fully operational water purification testbed for research in the design of secure cyber-physical systems. It produces five gallons/minute of doubly filtered water. A detailed account of the water purification process and some salient features and design considerations of SWaT can be found on our prior work \cite{eyasu_cybericps}. Concurrently, SWaT is based on closed-source and proprietary Allen Bradely PLCs. Hence, it is not possible to directly modify the firmware of these PLCs and to enforce memory-safety solutions. To alleviate this problem in our experimental evaluation, an open platform, named Open-SWaT, was designed. 

\begin{figure}
	\centering
  \includegraphics[scale=0.42]{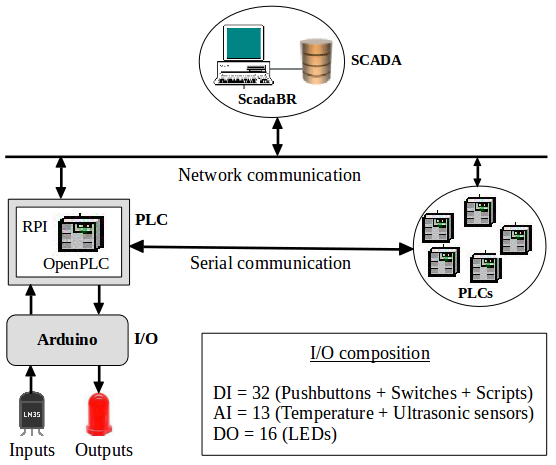} 
	\caption{Architecture of Open-SWaT}
	\label{fig:openswat}
\end{figure}

Open-SWaT  \cite{eyasu_cybericps,eyasu_essos} is 
a mini CPS based on open source PLCs that mimics the features and operational behaviors of SWaT. The PLCs are designed using OpenPLC \cite{openplc} -- an open source PLC that runs on top of Linux on Raspberry PI (RPI). 
With Open-SWaT, we reproduce operational profiles and details of SWaT. In particular, we reproduce the main factors 
that have substantial effect on the scan time and MSO of PLCs 
such as hardware specifications of PLCs (e.g. 200MHz of CPU speed and 2MB of user memory), a Remote Input/Output (RIO) terminal (containing 32 digital inputs (DI), 13 analog inputs (AI) and 16 digital outputs (DO)), real-time constraints (i.e. cycle time of PLCs), a PLC program (containing 129 instructions of several types), communication frequencies and a full SCADA system. The main purpose of designing Open-SWaT was to employ our CIMA approach on a realistic CPS. A high-level architecture of Open-SWaT is shown in Figure \ref{fig:openswat}. 
Due to space limitation, interested readers are referred to a detailed account of Open-SWaT in our prior papers \cite{eyasu_cybericps,eyasu_essos}.

\subsection{SecUTS}\label{subsec:secuts}
The Secure Urban Transportation System (SecUTS) is a CPS testbed designed 
to research on the security of a Metro SCADA 
system. The Metro SCADA system \cite{secuts_paper} 
comprises an {\it Integrated Supervisory Control System (ISCS)} and a {\it train signaling system}. ISCS 
integrates localized and centralized control and supervision of mechanical and electrical subsystems located at remote tunnels, depots, power substations and passenger stations. The entire Metro system can be remotely communicated, monitored, and controlled from the operation control center via the communication \REM{backbone}network. On the other hand, the signaling system facilitates communications between train-borne \REM{controllers}and track-side controllers. 
It also controls track-side equipments 
and train position localization. Modbus is used as a communication protocol among the devices in the ISCS. 
A detailed account of the Metro SCADA can be found on \cite{secuts_paper}. 

The SecUTS testbed provides facilities to examine several types of cyber attacks, such as message replay, forged message and memory-safety attacks, in the ISCS system and \REM{evaluate feasible} enforce proper countermeasures \REM{to prevent}against such attacks. 
However, the SecUTS testbed is also based on closed-source proprietary Siemens PLCs, hence we cannot directly enforce CIMA to these PLCs to detect and mitigate memory-safety attacks. Consequently, we similarly designed Open-SecUTS testbed (by mimicking SecUTS) using OpenPLC controller. It consists of 6 DI (emergency and control buttons) and 9 DO (tunnel and station lightings, ventilation and alarms). Subsequently, we enforced CIMA to Open-SecUTS and evaluated its practical applicability in a Metro SCADA system (See Section \ref{sec:evaluation_and_discussion}).

\subsection{Measuring runtime overheads} 

To measure the scan time and and compute memory-safety overheads of the PLCs in Open-SWaT and Open-SecUTS, a function is implemented using POSIX clocks (in nanosecond resolution). 
The function measures the execution time of each operation in the PLC scan cycle. Results will be then exported to external files for further manipulation, e.g., computing MSO and plotting graphs. We run 50,000 scan cycles for each PLC operation to measure the overall performance of the PLC. 

\section{Evaluation}\label{sec:evaluation_and_discussion}

This section discusses a detailed evaluation of our CIMA approach on Open-SWaT and Open-SecUTS. Subsequently, we discuss the experimental results to figure out whether our proposed approach is accurate enough to detect and mitigate memory-safety violations. We also discuss the efficiency of our approach in the context of CPS environment. 
In brief, we evaluate the proposed approach along four dimensions: \emph{1) Security guarantees} -- detection and mitigation accuracy of ASan and CIMA, respectively, \emph{2) Performance} -- tolerability of the runtime overhead of the proposed security measure in CPS environment, \emph{3) Resilience} -- its capability to ensure system availability and maintain physical-state resiliency in CPS even in the presence of memory-safety attacks, and \emph{4) its Memory usage overheads}. 

\subsection{Security guarantees}\label{subsec:security_evaluation}


To stress test the accuracy of our approach, we have evaluated CIMA against a wide-range of memory-safety vulnerabilities. This is to explore the accuracy of mitigating memory-safety vulnerabilities by our CIMA approach. As our CIMA approach is built on top of ASan, it is crucial that ASan {\em detects} a wide-range of memory-safety vulnerabilities accurately. According to the original results published for ASan~\cite{asan}, it detects memory-safety violations with high-accuracy -- without false positives for vulnerabilities such as (stack, heap and global) buffer under/overflows, use-after-free errors (dangling pointers), use-after-return errors, initialization order bugs and memory leaks . 
Only rare false negatives may appear for global buffer overflow  and use-after-free vulnerabilities due to some exceptions~\cite{asan}. 

CIMA effectively {\em mitigates} memory-safety violations, given that such a violation is detected by ASan at runtime. Therefore, the mitigation accuracy of our CIMA approach is exactly the same as the detection accuracy of ASan. 

As discussed in detail on Section \ref{sec:methodology}, we discovered two global buffer overflow vulnerabilities in the OpenPLC firmware. Both these vulnerabilities were successfully mitigated by our CIMA approach. Besides, throughout our evaluation, we did not discover any false positives or negatives in mitigating all the memory-safety violations detected by ASan.

\begin{table*}[htb]
\centering
\caption{Memory-safety overheads for the Open-SWaT Testbed}
\label{table:swat_mso}
{\small
\resizebox{\textwidth}{!}{%
\begin{tabular}{|l|l|l|l|l|l|l|l|l|l|l|l|l|l|}
\hline
\multirow{2}{*}{Operations} & \multirow{2}{*}{\begin{tabular}[c]{@{}l@{}}Number \\ of cycles\end{tabular}} & \multirow{2}{*}{\begin{tabular}[c]{@{}l@{}}Network\\ devices\end{tabular}} & \multirow{2}{*}{\begin{tabular}[c]{@{}l@{}}CPU speed\\ (in MHz)\end{tabular}} & \multicolumn{2}{c|}{Original ($T_{s}$)}                                                                        & \multicolumn{4}{c|}{ASan}                                                                                                                                                                                      & \multicolumn{4}{c|}{ASan + CIMA ($\hat{T_{s}}$)}                                                                                                                                                                              \\ \cline{5-14} 
                            &                                                                              &                                                                            &                                                                               & \begin{tabular}[c]{@{}l@{}}Mean\\ (in µs)\end{tabular} & \begin{tabular}[c]{@{}l@{}}Max\\ (in µs)\end{tabular} & \begin{tabular}[c]{@{}l@{}}Mean\\ (in µs)\end{tabular} & \begin{tabular}[c]{@{}l@{}}Max\\ (in µs)\end{tabular} & \begin{tabular}[c]{@{}l@{}}MSO\\ (in µs)\end{tabular} & \begin{tabular}[c]{@{}l@{}}MSO\\ (in \%)\end{tabular} & \begin{tabular}[c]{@{}l@{}}Mean\\ (in µs)\end{tabular} & \begin{tabular}[c]{@{}l@{}}Max\\ (in µs)\end{tabular} & \begin{tabular}[c]{@{}l@{}}MSO\\ (in µs)\end{tabular} & \begin{tabular}[c]{@{}l@{}}MSO\\ (in \%)\end{tabular} \\ \hline
                            \hline
Input scan                  & 50000                                                                        & 6                                                                          & 200                                                                           & 59.38                                                  & 788.12                                                & 118.44                                                 & 1132.32                                               & 59.09                                                 & 99.46                                                 & 122.86                                                 & 1151.35                                               & 63.48                                                 & 106.9                                                 \\ \hline
Program exec.               & 50000                                                                        & 6                                                                          & 200                                                                           & 69.09                                                  & 611.82                                                & 115.88                                                 & 720.36                                                & 46.79                                                 & 67.72                                                 & 118.97                                                 & 802.18                                                & 49.88                                                 & 72.2                                                  \\ \hline
Output update               & 50000                                                                        & 6                                                                          & 200                                                                           & 145.01                                                 & 981.09                                                & 185.37                                                 & 1125.45                                               & 40.36                                                 & 27.83                                                 & 199.89                                                 & 1213.62                                               & 54.88                                                 & 37.85                                                 \\ \hline
Full scan time              & 50000                                                                        & 6                                                                          & 200                                                                           & 273.48                                                 & 2381.03                                               & 419.69                                                 & 2978.13                                               & 146.21                                                & 53.46                                                 & 441.72                                                 & 3167.15                                               & 168.24                                                & 61.52                                                 \\ \hline
\end{tabular}}
}
\end{table*}

\begin{table*}[htb]
\centering
\caption{Memory-safety overheads for the Open-SecUTS Testbed}
\label{table:secuts_mso}
{\small
\resizebox{\textwidth}{!}{%
\begin{tabular}{|l|l|l|l|l|l|l|l|l|l|l|l|l|l|}
\hline
\multirow{2}{*}{Operations} & \multirow{2}{*}{\begin{tabular}[c]{@{}l@{}}Number \\ of cycles\end{tabular}} & \multirow{2}{*}{\begin{tabular}[c]{@{}l@{}}Network\\ devices\end{tabular}} & \multirow{2}{*}{\begin{tabular}[c]{@{}l@{}}CPU speed\\ (in MHz)\end{tabular}} & \multicolumn{2}{c|}{Original ($T_{s}$)}                                                                        & \multicolumn{4}{c|}{ASan}                                                                                                                                                                                      & \multicolumn{4}{c|}{ASan + CIMA ($\hat{T_{s}}$)}                                                                                                                                                                              \\ \cline{5-14} 
                            &                                                                              &                                                                            &                                                                               & \begin{tabular}[c]{@{}l@{}}Mean\\ (in µs)\end{tabular} & \begin{tabular}[c]{@{}l@{}}Max\\ (in µs)\end{tabular} & \begin{tabular}[c]{@{}l@{}}Mean\\ (in µs)\end{tabular} & \begin{tabular}[c]{@{}l@{}}Max\\ (in µs)\end{tabular} & \begin{tabular}[c]{@{}l@{}}MSO\\ (in µs)\end{tabular} & \begin{tabular}[c]{@{}l@{}}MSO\\ (in \%)\end{tabular} & \begin{tabular}[c]{@{}l@{}}Mean\\ (in µs)\end{tabular} & \begin{tabular}[c]{@{}l@{}}Max\\ (in µs)\end{tabular} & \begin{tabular}[c]{@{}l@{}}MSO\\ (in µs)\end{tabular} & \begin{tabular}[c]{@{}l@{}}MSO\\ (in \%)\end{tabular} \\ \hline
                            \hline
Input scan                  & 50000                                                                        & 1                                                                          & 200                                                                           & 59.84                                                  & 739.94                                                & 114.88                                                 & 902.01                                                & 55.04                                                 & 91.98                                                 & 115.07                                                 & 906.09                                                & 55.23                                                 & 92.3                                                  \\ \hline
Program exec.               & 50000                                                                        & 1                                                                          & 200                                                                           & 48.56                                                  & 488.38                                                & 91.36                                                  & 443.61                                                & 42.8                                                  & 88.14                                                 & 104.41                                                 & 676.19                                                & 55.85                                                 & 115.01                                                \\ \hline
Output update               & 50000                                                                        & 1                                                                          & 200                                                                           & 145.47                                                 & 850.62                                                & 175.59                                                 & 1045.34                                               & 30.12                                                 & 20.71                                                 & 178.91                                                 & 924.11                                                & 33.44                                                 & 22.99                                                 \\ \hline
Full scan time              & 50000                                                                        & 1                                                                          & 200                                                                           & 253.87                                                 & 2078.94                                               & 381.83                                                 & 2390.96                                               & 127.96                                                & 50.4                                                  & 398.39                                                 & 2506.39                                               & 144.52                                                & 56.93                                                 \\ \hline
\end{tabular}
}
}
\end{table*}

\subsection{Performance}\label{subsec:performance_evaluation}

According to the original article published for ASan~\cite{asan}, the average \REM{runtime overhead}MSO of ASan is 73\%. However, all measurements were taken on benchmarks different from ours and more importantly, in a non-CPS environment. With our CPS environment integrated in the Open-SWaT and Open-SecUTS, the average overhead induced by ASan is 53.46\% and 50.4\%, respectively. Additionally, our proposed CIMA approach induces 8.06\% and 6.53\% runtime overheads on Open-SWaT and Open-SecUTS, respectively. Thus, the overall runtime overhead of our security measure is 61.52\% (for Open-SWaT) and 56.93\% (for Open-SecUTS). A more detailed performance report, including the performance overhead of each PLC operation in both testbeds, is illustrated on Table~\ref{table:swat_mso} and \ref{table:secuts_mso}.

It is crucial to check whether the induced overhead by 
ASan and CIMA ($\hat{T}_{s}$) is tolerable in a CPS environment. To this end, we evaluate if this overhead respects the \emph{real-time constraints} of SWaT and SecUTS. For instance, consider the tolerability in average-case scenario. We observe that our proposed approach satisfies the condition of tolerability, as defined in Eq.~(\ref{eq:average-case-tolerability}). In particular, from Table \ref{table:swat_mso}, $mean(\hat{T}_{s}) = 441.72$µs, and $T_{c} = 10$ms; and from Table \ref{table:secuts_mso}, $mean(\hat{T}_{s}) = 398.39$µs, and $T_{c} = 150$ms. Consequently, Eq.~(\ref{eq:average-case-tolerability}) is satisfied and the overhead induced by our CIMA approach is both tolerable in SWaT and SecUTS. 


Similarly, considering the worst-case scenario, we evaluate if Eq.~(\ref{eq:worst-case-tolerability}) is satisfied. From Table \ref{table:swat_mso}, $max(\hat{T}_{s}) = 3167.15$µs, and $T_{c} =  10$ms; and from Table \ref{table:secuts_mso}, $max(\hat{T}_{s}) = 2506.39$µs, and $T_{c} = 150$ms. It is still tolerable, thus the proposed security measure satisfies SWaT's and SecUTS's real-time constraints 
in both scenarios. Therefore, despite high security guarantees provided by CIMA, its overhead is still tolerable in a CPS environment.

\begin{center}
\begin{tcolorbox}[width=\columnwidth, colback=white!25,arc=0pt,auto outer arc]
\textbf{CIMA and ASan together induce a runtime overhead of 61.52\% in Open-SWaT and 56.93\% in Open-SecUTS, whereas the overhead due to CIMA is only 8.06\% and 6.53\%, respectively. Despite this overhead, our proposed approach meets the hard real-time constraints for SWaT and SecUTS both in the average- and worst-case scenarios.}
\end{tcolorbox}
\end{center}

 \begin{table}[htb]
 \caption{\small Memory usage overheads for the Open-SWaT Testbed}
 \label{table:swat_memory_usage}
 \small
 \begin{center}
 
 \scalebox{0.95}{

\begin{tabular}{ |l|l|*{2}{c|}*{2}{c|}}
 \hline
Category &Original   & \multicolumn{2}{|c|}{ASan} & \multicolumn{2}{|c|}{ASan+CIMA}   
 \\
 \cline{3-6}
 &   &  Original & Overhead & Original & Overhead \\
 \hline
 \hline
 \shortstack {Virtual\\ Memory} & 62.97MB & 549.38MB & 8.72$\times$ & 557.5MB & 8.85$\times$
 \\
 \hline
 \shortstack {Real \\ Memory} & 8.17MB & 10.31MB & 1.26$\times$ & 11.2MB  & 1.37$\times$
 \\
 \hline
 \shortstack {Binary} & 144KB & 316KB & 2.19$\times$ & 324KB & 2.25$\times$
 \\
 \hline
 \shortstack {Shared \\ library } & 3196KB & 4288KB & 1.34$\times$ & 4288KB & 1.34$\times$
 \\
 \hline
 \end{tabular}
  }
  \end{center}
 \end{table}

 \begin{table}[htb]
 \caption{\small Memory usage overheads for the Open-SecUTS Testbed}
 \label{table:secuts_memory_usage}
 \small
 \begin{center}
 
 \scalebox{0.95}{

\begin{tabular}{ |l|l|*{2}{c|}*{2}{c|}}
 \hline
Category &Original   & \multicolumn{2}{|c|}{ASan} & \multicolumn{2}{|c|}{ASan+CIMA}   
 \\
 \cline{3-6}
 &   &  Original & Overhead & Original & Overhead \\
 \hline
 \hline
 \shortstack {Virtual\\ Memory} & 56.37MB & 489.29MB & 8.68$\times$& 490.6MB & 8.70$\times$
 \\
 \hline
 \shortstack {Real \\ Memory} & 8.76MB & 9.81MB & 1.12$\times$ & 10.21MB  & 1.17$\times$
 \\
 \hline
 \shortstack {Binary} & 136KB & 288KB & 2.12$\times$ & 296KB & 2.18$\times$
 \\
 \hline
 \shortstack {Shared \\ library } & 3196KB & 4288KB & 1.34$\times$ & 4288KB & 1.34$\times$
 \\
 \hline
 \end{tabular}
 }
 \end{center}
 \end{table}

\subsection{Resilience}\label{subsec:resilience}
One of the main contributions of our work is to empirically show the resilience of our CIMA approach against memory-safety attacks. Here, we evaluate how our mitigation strategy ensures \emph{availability} and \emph{physical-state resiliency} of a real-world CPS. As discussed in the preceding sections, CIMA does not render system unavailability. This is because it does not abort or restart the PLC when mitigating memory-safety attacks. In such a fashion, the availability of PLCs is ensured by our approach. 
As discussed in Section~\ref{modeling_physical_state_resiliency}, physical-state resiliency of a CPS can be affected by the memory-safety overhead (when the overhead is not tolerable due to the real-time constraint of the PLC) or the downtime of the PLC (when the PLC is unavailable for some reason). 

In the preceding section, we show that our CIMA approach ensures the memory-safety overhead to be tolerable. Hence, the additional overhead induced by CIMA does not affect the physical-state resiliency. Added to the fact is that the availability of SWaT and SecUTS is also ensured by CIMA via its very construction, as CIMA never aborts the system or leads to PLC downtime. Given that our proposed solution ensures availability of the PLC and also maintains the physical-state resiliency, we ensure the resilience of SWaT even in the presence of memory-safety attacks.

\begin{center}
\begin{tcolorbox}[width=\columnwidth, colback=white!25,arc=0pt,auto outer arc]
\textbf{CIMA ensures physical-state resiliency of SWaT and SecUTS, 
as $\hat{T}_s \leq T_c$ (cf. Eq.~(\ref{eq:downtime}) and Eq.~(\ref{eq:cls1})) holds in all of our experiments. This makes CIMA to be a security solution that ensures the SWaT and SecUTS systems to remain resilient even in the presence of memory-safety attacks.}
\end{tcolorbox}
\end{center}


\subsection{Memory usage overheads}
Finally, we evaluated the memory usage overheads of our CIMA approach. Tables \ref{table:swat_memory_usage} and \ref{table:secuts_memory_usage} summarize the increased virtual memory usage, real memory usage, binary size and shared library usage for the Open-SWaT and Open-SecUTS testbeds, respectively. The reported statistics are collected by reading \texttt{VmPeak}, \texttt{VmRSS}, \texttt{VmExe} and \texttt{VmLib} fields, respectively, from \emph{/proc/self/status}. In general, we observe a significant increase in virtual memory usages (8.85$\times$ for Open-SWaT and 8.70$\times$ for Open-SecUTS). This is primarily because of the allocation of large redzones with $malloc$ (as part of the ASan approach).  However, the real memory usage overhead is only 1.37$\times$ (for Open-SWaT) and 1.17$\times$ for Open-SecUTS. We believe these overheads are still acceptable since most PLCs nowadays come with at least 1GB memory size. Moreover, the increased memory size is an acceptable trade-off in the light of strong mitigation mechanics provided by our CIMA approach. Finally, we observe that CIMA introduces 
negligible 
memory usage overhead over ASan, meaning the majority of memory-usage overhead is attributed to the usage of ASan. 

\section{Related work}\label{sec:related_work}

CIMA is built on top of ASan. ASan~\cite{asan} is a fast memory-safety tool based on code-instrumentation. It covers a wide range of temporal memory errors, such as use-after-free, use-after-return and memory leaks, and spatial memory errors such as stack, heap and global buffer overflows. 
It is also a standard tool included in the GCC and LLVM compiler infrastructures as a \REM{default}memory-safety checker. However, its mitigation approach (in normal mode) is to simply abort the program whenever a memory-safety violation is detected. 
This is not acceptable in most critical systems, such as CPS and ICS, with stringent time constraints. In such systems,  availability is of the utmost importance. This makes ASan to be impractical for critical systems with hard real-time constraints. 
ASan does also provide a special mode to continue execution even after detecting a memory-safety error. However, this mode does not provide any protection against memory-safety attacks. 

Softbound~\cite{softbound} and its extension CETS~\cite{cets} guarantee a complete memory-safety. However, such guarantees arrive with the cost of a very high runtime overhead (116\%). Such a high performance overhead is unlikely to be tolerable due to the real-time constraints imposed on a typical CPS. Moreover, Softbound and CETS do not implement a mitigation strategy to consider the physical-state resiliency in a CPS environment.    

SafeDispatch~\cite{safedispatch} is a fast memory-safety tool developed within the LLVM infrastructure. SafeDispatch also involves exhaustive performance optimizations to make the overhead just 2.1\%. However, SafeDispatch is not supported by an  appropriate mitigation strategy that guarantees system availability in the presence of memory-safety attacks. 
Thus, its applicability to the CPS environment is limited. 

Sting~\cite{sting_song,vsef_song} is an end-to-end self-healing architecture developed against memory-safety attacks. It detects attacks with address space layout randomization (ASLR) and system-call-based anomaly detection techniques. However, ASLR can be defeated by code-reuse attacks and system-call-based detections (e.g. control-flow integrity) can be bypassed by data injection attacks. To diagnosis the root cause of the attack, Sting leverages a heavy-weight static taint analysis. Furthermore, it performs periodic check-pointing and continuously records system calls to resume the victim program from an earlier safe state. These techniques bring significant performance and memory usage overhead. As reported~\cite{sting_song}, there are cases where the corrupted program cannot be recovered and requires to be restarted. Therefore, the performance and memory usage overheads and the system unavailability problems limit the applicability of Sting to a CPS environment. 

ROPocop~\cite{ropocop} is a dynamic binary code instrumentation framework against code injection and code reuse attacks. It relies on Windows x86 binaries to detect such attacks. Its runtime overhead is 240\%, which is significantly high and is unlikely be tolerable due to the real-time constraints in CPS. Moreover, it is unclear what kind of mitigation strategies were 
incorporated with ROPocop.   

Over the past decades, a number of control-flow integrity (CFI) based solutions (e.g., \cite{cfi_cots,cfi_gcc,cfi_sp,cfi_rockjit}) have been developed to defend against memory-safety attacks. The main objective behind these solutions is ensuring the control-flow integrity of a program. Therefore, they aim to prevent attacks from redirecting the flow of program execution. These solutions also offer a slight performance advantage over other countermeasures, such as code-instrumentation based countermeasures. However, CFI-based solutions generally have the following limitations: 
{\em (i)} determining the required control-flow graph (often using static analysis) is difficult and requires a substantial amount of memory; {\em (ii)} attacks that are not diverting the control-flow of the program cannot be detected (e.g. data oriented attacks~\cite{cfi_data-attacks}); {\em (iii)} finally, these solutions are mainly to detect memory-safety attacks, but do not implement mitigation strategies against the attacks. Consequently, the applicability of CFI-based solutions is limited in a CPS environment.


In summary, to the best of our knowledge, there is no prior work that efficiently detects and mitigates memory-safety attacks without compromising the real-time constraints or availability of the underlying system. Moreover, we are not aware of any work that designs and evaluates memory-safety attack mitigation techniques in the light of real-time constraints and physical-state resiliency imposed on critical systems. In this paper, CIMA  bridges this gap of research. 


\section{Threats to validity}
\label{sec:threats}

The effectiveness of CIMA critically depends on the following: 

\begin{enumerate}


\item \emph{\bf Uncovered memory-safety errors}: ASan does not cover some memory-safety errors such as uninitialized memory reads and some use-after-return bugs. Thus, CIMA does not also mitigate these errors. However, these errors are less critical and unlikely to be exploited in practice. Moreover, the central idea behind CIMA is unaffected by such limitation. Specifically, CIMA leverages ASan as an off-the-shelf capability to detect the memory-safety errors. Any improvement on memory-safety attack detection tool will also automatically improve the coverage of attacks mitigated via CIMA.  

\item \emph{\bf Experimental limitations}: We conducted our experiments only on a water treatment and Metro SCADA systems. Nevertheless, our proposed approach is generic and it can be applied on different CPSs with various real-time constraints. In the future, we intend to conduct further experiments on power grid and robotics systems.  
\item \emph{\bf Limitations due to proprietary PLC code}: SWaT and SecUTS are realistic CPS \REM{water treatment}testbeds \REM{that contains}containing a set of real-world vendor-supplied PLCs. However, their PLCs are proprietary and closed-source. Hence, we were unable to incorporate our security solution directly to these PLCs. Instead, we designed our testbeds (i.e. Open-SWaT and Open-SecUTS) by mimicking all operational details of SWaT and SecUTS, respectively. While designing the testbeds, 
we have paid careful attention to all design features and constraints imposed on SWaT and SecUTS, especially timing-related constraints. Hence, we can provide high confidence on our results obtained from Open-SWaT and Open-SecUTS to be applicable also to the real SWaT and SecUTS systems. 

\end{enumerate}

\section{Conclusion}\label{sec:conclusion}

In this paper, we propose CIMA, a resilient mitigation strategy to ensure protection against a wide variety of memory-safety attacks. The main advantage of CIMA is that it mitigates memory-safety attacks in a time-critical environment and ensures the system availability by skipping only the instructions that exhibit illegal memory accesses. Such an advantage makes CIMA to be an attractive choice of security measure for cyber-physical systems and critical infrastructures, which often impose strict timing constraints. To this end, we evaluate our approach on a real-world CPS testbed. Our evaluation reveals that CIMA mitigates memory-safety errors with acceptable runtime and memory-usage overheads. Moreover, it ensures that the resiliency of the physical states are maintained despite the presence of memory-safety attacks. Although we evaluated our approach only in the context of CPS, CIMA is also applicable and useful for any system under the threat of memory-safety attacks. 

From conceptual point of view, CIMA provides a fresh outlook over mitigating memory-safety attacks. In future, we plan to build upon our approach to understand the value of CIMA across a variety of systems beyond CPS. We also plan to leverage our CIMA approach for live patching. In particular, at its current state, CIMA does not automatically fix the memory-safety vulnerabilities of the victim code (although CIMA does ensure that the vulnerabilities are not exploited at runtime). Instead, it generates a report for the developer to help produce a patched version of the code. In future, we will use the report generated by CIMA to automatically identify the pattern of illegal memory accesses and fix the code accordingly. 

\bibliographystyle{splncs}
\bibliography{mybib}

\begin{thebibliography}{10}

\bibitem{sok}
Szekeres, L., Payer, M., Wei, T., Song, D.:
\newblock Sok: Eternal war in memory.
\newblock 2013 IEEE Symposium on Security and Privacy (2013)

\bibitem{memory_safety_attacks_survey}
Saito, T., Watanabe, R., Kondo, S., Sugawara, S., Yokoyama, M.:
\newblock A survey of prevention/mitigation against memory corruption attacks.
\newblock 2016 19th International Conference on Network-Based Information
  Systems (NBiS) (2016)

\bibitem{code_injection_survey}
Younan, Y., Joosen, W., Piessens, F.:
\newblock Code injection in c and c++ : A survey of vulnerabilities and
  countermeasures.
\newblock Technical report (2004)

\bibitem{code_injection_attacks}
Francillon, A., Castelluccia, C.:
\newblock Code injection attacks on harvard-architecture devices.
\newblock In: Proceedings of the 15th ACM Conference on Computer and
  Communications Security (CCS'08). (2008)

\bibitem{code_reuse_attacks1}
Snow, K.Z., Monrose, F., Davi, L., Dmitrienko, A., Liebchen, C., Sadeghi, A.:
\newblock Just-in-time code reuse: On the effectiveness of fine-grained address
  space layout randomization.
\newblock In: Proceedings of the IEEE Symposium on Security and Privacy
  (SP'13), Washington, USA (2013)

\bibitem{code_reuse_attacks2}
Dahse, J., Krein, N., Holz, T.:
\newblock Code reuse attacks in php: Automated pop chain generation.
\newblock In: Proceedings of the ACM SIGSAC Conference on Computer and
  Communications Security (CCS'14). (2014)

\bibitem{brop}
Bittau, A., Belay, A., Mashtizadeh, A., Mazi\'{e}res, D., Boneh, D.:
\newblock Hacking blind.
\newblock In: Proceedings of the 2014 IEEE Symposium on Security and Privacy.
  SP'14 (2014)  227 -- 242

\bibitem{eyasu_cybericps}
Chekole, E.G., Castellanos, J.H., Ochoa, M., Yau, D.K.Y.:
\newblock Enforcing memory safety in cyber-physical systems.
\newblock In: Katsikas S. et al. (eds) Computer Security. SECPRE 2017,
  CyberICPS 2017

\bibitem{eyasu_essos}
Chekole, E.G., Chattopadhyay, S., Ochoa, M., Huaqun, G.:
\newblock Enforcing full-stack memory safety in cyber-physical systems.
\newblock In: Proceedings of the International Symposium on Engineering Secure
  Software and Systems (ESSoS'18), Springer International Publishing (2018)

\bibitem{asan}
Serebryany, K., Bruening, D., Potapenko, A., Vyukov, D.:
\newblock Addresssanitizer: a fast address sanity checker.
\newblock In: Proceedings of the USENIX conference on Annual Technical
  Conference (USENIX ATC'12). (2012)

\bibitem{safedispatch}
Jang, D., Tatlock, Z., Lerner, S.:
\newblock Safedispatch: Securing c virtual calls from memory corruption
  attacks.
\newblock Proceedings 2014 Network and Distributed System Security Symposium
  (2014)

\bibitem{ropocop}
Follner, A., Bodden, E.:
\newblock Ropocop — dynamic mitigation of code-reuse attacks.
\newblock Journal of Information Security and Applications (2016)

\bibitem{cfi_cots}
Zhang, M., Sekar, R.:
\newblock Control flow integrity for cots binaries.
\newblock In: Proceedings of the USENIX Security Symposium (USENIX'13). (2013)

\bibitem{cfi_gcc}
Tice, C., Roeder, T., Collingbourne, P., Checkoway, S., Erlingsson, {\'U}.,
  Lozano, L., Pike, G.:
\newblock Enforcing forward-edge control-flow integrity in gcc \& llvm.
\newblock In: Proceedings of the 23rd USENIX Security Symposium. USENIX'14
  (2014)  941--955

\bibitem{cfi_sp}
Ge, X., Talele, N., Payer, M., Jaeger, T.:
\newblock Fine-grained control-flow integrity for kernel software.
\newblock In: 2016 IEEE European Symposium on Security and Privacy. (2016)

\bibitem{softbound}
Nagarakatte, S., Zhao, J., Martin, M.M., Zdancewic, S.:
\newblock Softbound: Highly compatible and complete spatial memory safety for
  {C}.
\newblock In: Proceedings of the 2009 ACM SIGPLAN conference on Programming
  language design and implementation. PLDI'09 (2009)

\bibitem{cets}
Nagarakatte, S., Zhao, J., Martin, M.M., Zdancewic, S.:
\newblock Cets: Compiler enforced temporal safety for {C}.
\newblock In: Proceedings of the 2010 International Symposium on Memory
  Management (ISMM'10). (2010)

\bibitem{memsafe}
Simpson, M.S., Barua, R.K.:
\newblock Memsafe: Ensuring the spatial and temporal memory safety of c at
  runtime.
\newblock Software: Practice and Experience \textbf{43}(1) (2013)  93--128

\bibitem{drmemory}
Bruening, D., Zhao, Q.:
\newblock Practical memory checking with dr. memory.
\newblock In: Proceedings of the 9th Annual IEEE/ACM International Symposium on
  Code Generation and Optimization. CGO'11

\bibitem{ccured}
Necula, G.C., Condit, J., Harren, M., McPeak, S., Weimer, W.:
\newblock Ccured: Type-safe retrofitting of legacy software.
\newblock ACM Trans. Program. Lang. Syst. \textbf{27}(3) (2005)

\bibitem{mudflap}
{F. Ch. Eigler}:
\newblock {Mudflap: pointer use checking for C/C++}.
\newblock In: GCC Developer's Summit, Red Hat Inc. (2003)

\bibitem{sting_song}
Newsome, J., Brumley, D., Song, D.:
\newblock Sting: An end-to-end self-healing system for defending against
  zero-day worm attacks on commodity software (2005)

\bibitem{vsef_song}
Newsome, J., Brumley, D., Song, D.:
\newblock Vulnerability-specific execution filtering for exploit prevention on
  commodity software.
\newblock In: Proceedings of the 13th Symposium on Network and Distributed
  System Security (NDSS'05). (2005)

\bibitem{vbs_song}
Brumley, D., Newsome, J., Song, D., Wang, H., Jha, S.:
\newblock Towards automatic generation of vulnerability-based signatures.
\newblock In: Proceedings of IEEE Symposium on Security and Privacy (SP'06),
  DC, USA (2006)

\bibitem{live_patching1}
Smirnov, A., Chiueh, T.:
\newblock Automatic patch generation for buffer overflow attacks.
\newblock Third International Symposium on Information Assurance and Security
  (2007)

\bibitem{cve_ab}
CVE-2016-5814.
\newblock \url{https://cve.mitre.org/cgi-bin/cvename.cgi?name=CVE-2016-5814}
  (2016)

\bibitem{cve_ab2}
CVE-2012-6438.
\newblock \url{https://cve.mitre.org/cgi-bin/cvename.cgi?name=CVE-2012-6438}
  (2012)

\bibitem{cve_ab3}
CVE-2012-6436.
\newblock \url{https://cve.mitre.org/cgi-bin/cvename.cgi?name=CVE-2012-6436}
  (2012)

\bibitem{cve_siemens1}
CVE-2013-0674.
\newblock \url{https://cve.mitre.org/cgi-bin/cvename.cgi?name=CVE-2013-0674}
  (2013)

\bibitem{cve_siemens2}
CVE-2015-1449.
\newblock \url{https://cve.mitre.org/cgi-bin/cvename.cgi?name=CVE-2015-1449}
  (2015)

\bibitem{cve_sem1}
CVE-2012-0929.
\newblock \url{https://cve.mitre.org/cgi-bin/cvename.cgi?name=CVE-2012-0929}
  (2012)

\bibitem{cve_sem2}
CVE-2015-7937.
\newblock \url{https://cve.mitre.org/cgi-bin/cvename.cgi?name=CVE-2015-7937}
  (2015)

\bibitem{cve_abb}
CVE-2011-5007.
\newblock \url{https://cve.mitre.org/cgi-bin/cvename.cgi?name=CVE-2011-5007}
  (2011)

\bibitem{asgithub}
github repository, A.:
\newblock Comparison of addresssanitizer with other memory safety tools.
\newblock
  \url{https://github.com/google/sanitizers/wiki/AddressSanitizerComparisonOfMemoryTools}
  (2015)

\bibitem{coop}
Schuster, F., Tendyck, T., Liebchen, C., Davi, L., Sadeghi, A.R., Holz, T.:
\newblock Counterfeit object-oriented programming: On the difficulty of
  preventing code reuse attacks in c applications.
\newblock 2015 IEEE Symposium on Security and Privacy (2015)

\bibitem{basic_block}
GCC:
\newblock Gcc basic blocks.
\newblock \url{https://gcc.gnu.org/onlinedocs/gccint/Basic-Blocks.html} (2018)

\bibitem{swat}
SWaT:
\newblock Secure water treatment (swat) testbed (2018)

\bibitem{openplc}
OpenPLC:
\newblock Openplc.
\newblock \url{http://www.openplcproject.com/} (2018)

\bibitem{secuts_paper}
Zhou, L., Guo, H., Li, D., Wong, J.W., Zhou, J.:
\newblock Mind the gap: Security analysis of metro platform screen door system.
\newblock In: Proceedings of the Singapore Cyber-Security RandD Conference
  (SG-CRC'17). (2017)

\bibitem{cfi_rockjit}
Niu, B., Tan, G.:
\newblock Rockjit: Securing just-in-time compilation using modular control-flow
  integrity.
\newblock In: Proceedings of the ACM SIGSAC Conference on Computer and
  Communications Security (CCS'14). (2014)

\bibitem{cfi_data-attacks}
Hu, H., Shinde, S., Adrian, S., Chua, Z.L., Saxena, P., Liang, Z.:
\newblock Data-oriented programming: On the expressiveness of non-control data
  attacks.
\newblock 2016 IEEE Symposium on Security and Privacy (2016)

\end{thebibliography}





%





\end{document}